\documentclass[aps,prl,twocolumn,superscriptaddress]{revtex4-2}

\usepackage{amssymb}
\usepackage{natbib}

\usepackage{multirow}
\usepackage{bm}
\usepackage{amsmath}
\usepackage{graphicx}
\usepackage{gensymb}
\usepackage{svg}
\usepackage{subfigure}
\usepackage{natbib}
\usepackage{amsfonts}
\usepackage{mathrsfs}
\usepackage{braket}
\usepackage{xcolor}
\usepackage{tabularx}
\usepackage[toc,page,title,titletoc,header]{appendix}
\usepackage[colorlinks,urlcolor=black, linkcolor=blue,citecolor=blue,anchorcolor=blue]{hyperref}
\usepackage{dsfont,amsthm,amsbsy}

\newcommand{\dudn}{\text{d}\mu/\text{d}n}

\newcommand{\cmmnt}[1]{\ignorespaces}

\usepackage{fancyhdr}
\usepackage{lineno}

\begin{document}

\title{Hofstadter states and reentrant charge order in a semiconductor moir\'e lattice}

\author{Carlos R. Kometter}
\thanks{These authors contributed equally}
\affiliation{Geballe Laboratory for Advanced Materials, Stanford, CA 94305, USA}
\affiliation{Department of Physics, Stanford University, Stanford, CA 94305, USA}

\author{Jiachen Yu}
\thanks{These authors contributed equally}
\affiliation{Geballe Laboratory for Advanced Materials, Stanford, CA 94305, USA}
\affiliation{Department of Applied Physics, Stanford University, Stanford, CA 94305, USA}

\author{Trithep Devakul}
\thanks{These authors contributed equally}
\affiliation{Department of Physics, Massachusetts Institute of Technology, Cambridge, Massachusetts 02139, USA}

\author{Aidan P. Reddy}
\affiliation{Department of Physics, Massachusetts Institute of Technology, Cambridge, Massachusetts 02139, USA}

\author{Yang Zhang}
\affiliation{Department of Physics, Massachusetts Institute of Technology, Cambridge, Massachusetts 02139, USA}

\author{Benjamin A. Foutty}
\affiliation{Geballe Laboratory for Advanced Materials, Stanford, CA 94305, USA}
\affiliation{Department of Physics, Stanford University, Stanford, CA 94305, USA}

\author{Kenji Watanabe}
\affiliation{Research Center for Functional Materials, National Institute for Material Science, 1-1 Namiki, Tsukuba 305-0044, Japan}

\author{Takashi Taniguchi}
\affiliation{International Center for Materials Nanoarchitectonics, National Institute for Material Science, 1-1 Namiki, Tsukuba 305-0044, Japan}

\author{Liang Fu}
\affiliation{Department of Physics, Massachusetts Institute of Technology, Cambridge, Massachusetts 02139, USA}

\author{Benjamin E. Feldman}
\email{bef@stanford.edu}
\affiliation{Geballe Laboratory for Advanced Materials, Stanford, CA 94305, USA}
\affiliation{Department of Physics, Stanford University, Stanford, CA 94305, USA}
\affiliation{Stanford Institute for Materials and Energy Sciences, SLAC National Accelerator Laboratory, Menlo Park, CA 94025, USA}


\begin{abstract}
The emergence of moir\'e materials with flat bands provides a platform to systematically investigate and precisely control correlated electronic phases. Here, we report local electronic compressibility measurements of a twisted WSe$_2$/MoSe$_2$ heterobilayer which reveal a rich phase diagram of interpenetrating Hofstadter states and electron solids. We show that this reflects the presence of both flat and dispersive moir\'e bands whose relative energies, and therefore occupations, are tuned by density and magnetic field. At low densities, competition between moir\'e bands leads to a transition from commensurate arrangements of singlets at doubly occupied sites to triplet configurations at high fields. Hofstadter states (i.e., Chern insulators) are generally favored at high densities as dispersive bands are populated, but are suppressed by an intervening region of reentrant charge-ordered states in which holes originating from multiple bands cooperatively crystallize. Our results reveal the key microscopic ingredients that favor distinct correlated ground states in semiconductor moir\'e systems, and they demonstrate an emergent lattice model system in which both interactions and band dispersion can be experimentally controlled.

\end{abstract}

\maketitle
In correlated many-body systems, two distinct classes of phases serve as the pillars of our microscopic understanding: crystalline, in which the constituent particles are strongly localized in a periodic configuration; and fluid, in which the wavefunctions of the particles are itinerant. The interplay and transition between these two regimes is of central interest in diverse contexts, ranging from metal-insulator transitions \cite{kravchenko_metalinsulator_2003} to two-dimensional electron gases in a quantizing magnetic field \cite{sarma_perspectives_2008} to ultracold quantum gases \cite{gross_quantum_2021}. These two types of phases, if they coexist in the same physical system, often compete due to the intrinsic incompatibility between their microscopic degrees of freedom. 

In recent years, moir\'e transition metal dichalcogenide (TMD) systems have emerged as a new manifestation of strongly correlated quantum matter because they can host flat electronic bands \cite{kennes_moire_2021,mak_semiconductor_2022}. In heterobilayers, the low-energy Hamiltonian takes a simple form when the interlayer tunneling is suppressed: the bare electronic structure of one constituent layer is subject to a long wavelength moir\'e potential imposed by the other. Electron (hole) Wannier orbitals are localized into minima (maxima) of the moir\'e potential, thereby forming a superlattice that can be well described by an extended Hubbard model \cite{wuHubbardModelPhysics2018,pan_quantum_2020,tangSimulationHubbardModel2020a,reganMottGeneralizedWigner2020,xuCorrelatedInsulatingStates2020a,liChargeorderenhancedCapacitanceSemiconductor2021,jin_stripe_2021,huang_correlated_2021,liu_excitonic_2021,chu_nanoscale_2020,li_imaging_2021,li_continuous_2021}. In the presence of long-range Coulomb interactions, these systems naturally favor commensurate charge-ordered states that maximally benefit from the underlying potential at densities corresponding to fractional fillings of electrons/holes per moir\'e unit cell, known as generalized Wigner crystals \cite{tangSimulationHubbardModel2020a,reganMottGeneralizedWigner2020,xuCorrelatedInsulatingStates2020a,liChargeorderenhancedCapacitanceSemiconductor2021,jin_stripe_2021,huang_correlated_2021,liu_excitonic_2021,chu_nanoscale_2020,li_imaging_2021,li_continuous_2021}. Electron solids are known to compete closely with (fractional) quantum Hall liquids in two-dimensional electron gases subject to high magnetic fields. However, whether this interplay exists in a lattice system, and how it is modified by the strong underlying periodic potential, is unknown. To date, study of quantum Hall analogues on a lattice, known as the Hofstadter butterfly, has been limited to graphene-based heterostructures whose moir\'e potential modulation is weak \cite{sarma_perspectives_2008}, and such states have not been reported in TMD systems. 

Here we report local electronic compressibility measurements of a twisted MoSe$_2$/WSe$_2$ heterobilayer. The small lattice mismatch between constituents leads to a long moir\'e wavelength at low twist angles and enables us to explore an unprecedentedly large range of filling and magnetic flux per moir\'e unit cell. We observe a complex pattern of competing/coexisting Hofstadter and charge-ordered states, with phase transitions separating regimes of distinct behavior. We show that this can be rationalized in terms of flat and dispersive moir\'e bands, with changes in occupancy tuned by both density and magnetic field. Within the flat bands, our measurements demonstrate multiple commensurate arrangements of doubly occupied moir\'e sites. These are generalized Wigner crystals for doublons, whose charge gaps exhibit nontrivial dependence on magnetic field. In the dispersive bands, we observe interpenetrating Hofstadter states at high magnetic flux, likely due to interaction-driven band reordering. Most strikingly, we find reentrance of charge-ordered states at moderate fields and densities, formed by collective crystallization of holes from both flat and dispersive bands. Our results establish semiconductor moir\'e superlattices as a rich platform to investigate phase transitions between crystalline and fluid correlated states, and they demonstrate that both the real space structure of the electronic orbitals and strong Coulomb interactions play essential roles in the transitions. 

\section{Coexistence of Hofstadter and charge-ordered states}
\begin{figure*}[t]
    \renewcommand{\thefigure}{\arabic{figure}}
    \centering
    \includegraphics[scale=1.0]{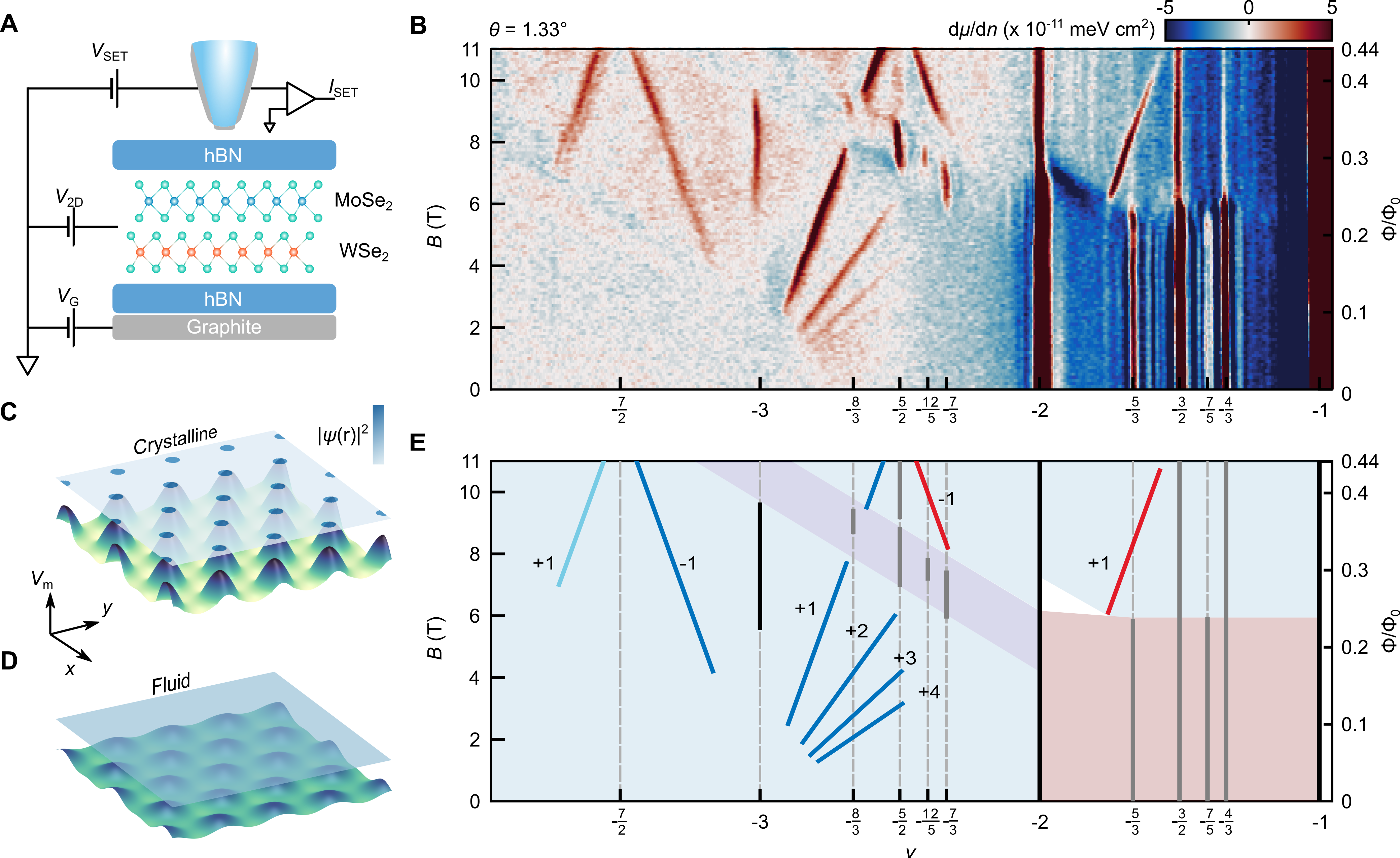}
    \caption{\textbf{Competing Hofstadter and charge-ordered states.} \textbf{(A)} Setup of the scanning single-electron transistor (SET) measurement. $V_{\textrm{2D}}$ and $V_{\textrm{G}}$ are voltages applied to the sample and back gate, respectively. $V_{\textrm{SET}}$ and $I_{\textrm{G}}$ are bias across and current through the SET (see Supplementary Material). \textbf{(B)} Local inverse electronic compressibility $\dudn$ measured at a location with twist angle $\theta=1.33^\circ$, as a function of magnetic field $B$ (with the number of flux quanta per moir\'e unit cell $\Phi/\Phi_0$ indicated on the right axis) and hole doping per moir\'e unit cell $\nu$. Data exceeding the limits of the color scale are truncated.  \textbf{(C} and \textbf{D)} Cartoon illustration of the two distinct types of ground states, crystalline (C) and fluid (D). A strong underlying moir\'e potential ($V_m$) favors spatially ordered charge arrangement (C), whereas Hartee screening from filled states weakens the effective potential and supports extended quantum liquid ground states (D). \textbf{(E)} Incompressible states identified from (B). Sloped linear trajectories correspond to Hofstadter states, whereas vertical trajectories at fractional $\nu$ correspond to charge-ordered states. The Chern numbers $t$ of the Hofstadter states are labeled, and their intercepts $s$ are denoted by their line colors: red ($s=-2$), blue ($s=-3$), and cyan ($s=-4$). The background colors reflect distinct regimes where different classes of states are favored.} 
    \label{fig:Fig1}
\end{figure*}

The experimental measurement setup is schematically illustrated in Fig. \ref{fig:Fig1}A. We use a scanning single-electron transistor (SET) to study an AA-stacked MoSe$_2$/WSe$_2$ heterobilayer with a small twist angle between layers. Figure \ref{fig:Fig1}B shows the local inverse electronic compressibility, $\dudn$, as a function of magnetic field $B$ and hole filling per moir\'e unit cell $\nu$ at a location with twist angle $\theta = 1.33^\circ$ (see Supplementary Material for details of twist angle determination). We observe an intricate pattern of incompressible states that follow linear trajectories in the $\nu-B$ plane. Each state can be described by two rational quantum numbers $(t, s)$ through the Diophantine equation $\nu=t(\Phi/\Phi_0)+s$, where $\Phi_0=h/e$ is the flux quantum, $t$ is the Chern number, and $s$ is the intercept at zero magnetic flux $\Phi$. We classify these gapped states into two distinct types. Gaps with $t=0$ are topologically trivial and generally correspond to charge-ordered (crystalline) insulators in which holes form localized moments at the moir\'e lattice sites (Fig. \ref{fig:Fig1}C). In addition to gapped states at integer fillings, we observe incompressible states with fractional $s$, which are generalized Wigner crystals. In contrast, gaps with nonzero (integer) $t$ emerge at finite $B$ and carry a nonzero Chern number. These are Hofstadter (Chern) insulators, topological states whose wavefunctions consist of moving wavepackets formed by magnetic Bloch states \cite{chang_berry_1996} (Fig. \ref{fig:Fig1}D).


The Hofstadter and charge-ordered states occur in distinct regions of the $\nu-B$ plane. At low magnetic fields, charge order is present at multiple commensurate filling factors between $-2<\nu<-1$: $-3/2, -4/3, -5/3$, and $-7/5$. In this range of filling, we observe an abrupt change at a critical field $B_c\approx6$ T, above which the charge-ordered states weaken and a Hofstadter state with $(t,s)=(1,-2)$ appears. The transition manifests over this whole range of filling and the critical field depends only weakly on density, suggesting that it is driven by a change in the underlying moir\'e bands.

Very different behavior occurs for $\nu<-2$, where an unexpected series of incompressible Hofstadter states instead emerge. Most of these states emanate from $s=-3$, with a singly-degenerate Landau fan at low field spanning towards lower hole doping, reminiscent of the asymmetric Landau fan in magic-angle twisted bilayer graphene \cite{cao_unconventional_2018}. At high fields, multiple Hofstadter states that respectively extrapolate to $s = -2, -3$ and $-4$ are visible. Remarkably, reentrant charge order is present at intermediate fields for $\nu<-2$, within a slanted narrow strip of width $\Delta\nu \approx 1/2$ that penetrates through and completely suppresses the Hofstadter states. The fractional fillings (modulo an integer) at which charge order occurs in this region are identical to those between $-2<\nu<-1$ at low fields. Together, these observations reveal a rich phase diagram of competing ground states with differing qualitative character. The incompressible states that we observe, and the regions of distinct phenomenology are summarized in Fig. \ref{fig:Fig1}E. 

\begin{figure*}[t]
    \renewcommand{\thefigure}{\arabic{figure}}
    \centering
    \includegraphics[scale =1.0]{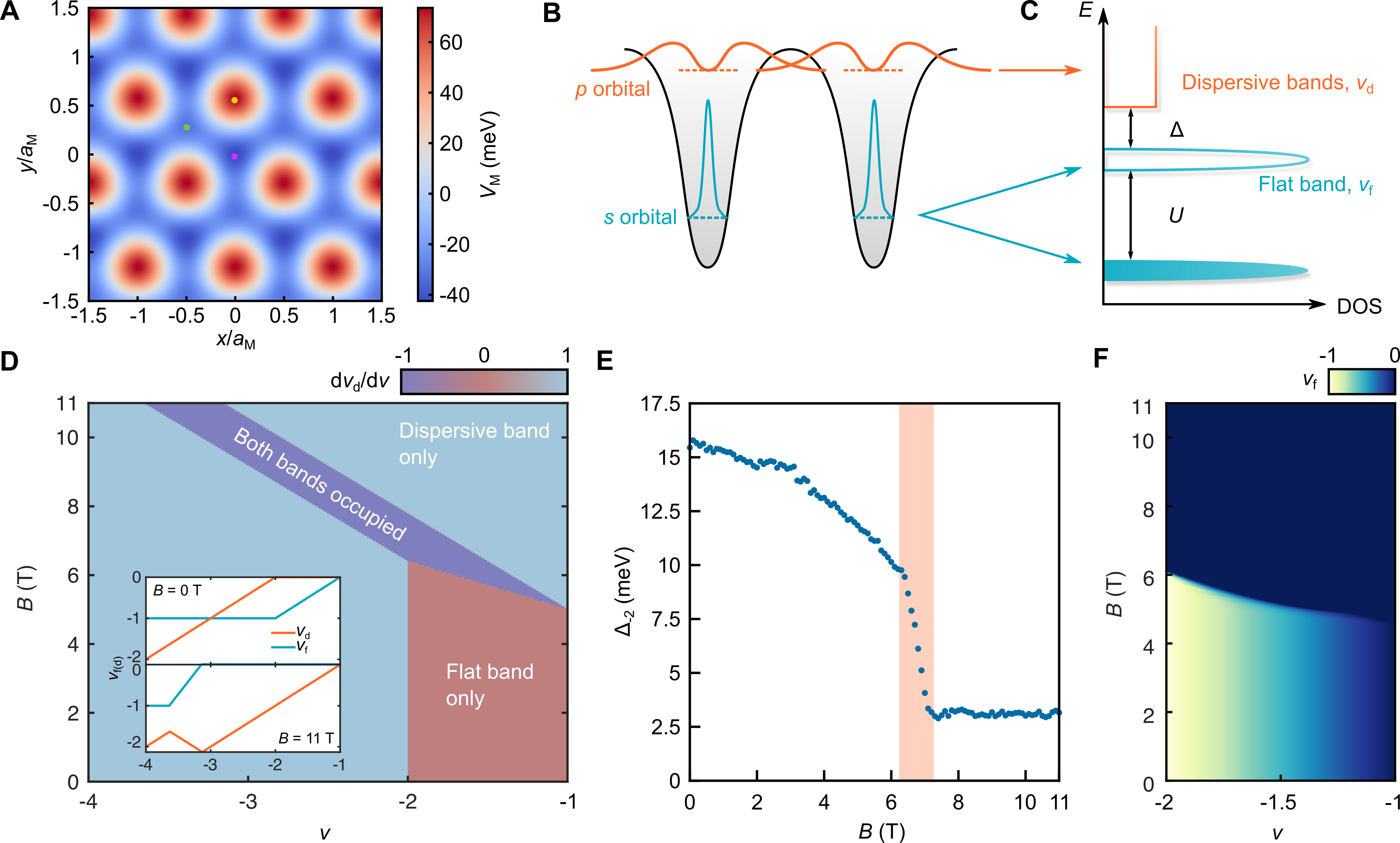}
    \caption{\textbf{Tuning between flat and dispersive moir\'e bands.} \textbf{(A)} Density function theory (DFT) calculation of the real space moir\'e potential profile. $V_M$, moir\'e potential. $a_M$, moir\'e lattice constant. Yellow, green and pink dots denote XM, MX and MM stacking configurations, respectively. \textbf{(B} and \textbf{C)} Schematic of eigenstates in the moir\'e potential wells (B) whose hybridization results in flat and dispersive moir\'e bands (C). On-site Coulomb interaction $U$ further splits the flat bands into lower and upper Hubbard bands (cyan), the latter of which has an energy offset $\Delta$ from the dispersive bands (orange). \textbf{(D)} Change of dispersive band occupation $d\nu_d/d\nu$, calculated within the phenomenological Stoner model. Inset, Upper Hubbard (dispersive) band filling factor $\nu_{f(d)}$ as a function of $\nu$ calculated at zero field (top) and at $B=11$ T (bottom). \textbf{(E)} Magnetic field dependence of the $\nu=-2$ gap size. The gap shrinks rapidly in the shaded region where the transition occurs. \textbf{(F)} $\nu_f$ as a function of $\nu$ and $B$ calculated within the Hubbard model. The rapid depletion of $\nu_f$ occurs at the gap closing field in (E).} 
    \label{fig:band-crossing}
\end{figure*}

In addition to the incompressible states, our measurement also reveals large areas of negative compressibility, indicative of flat bands and strong electron-electron interactions \cite{eisenstein_negative_1992,skinner_anomalously_2010,foutty_tunable_2022}. We observe a negative background over the entire range  $-2<\nu<-1$, with the most negative values near $\nu=-1$ and a weaker but qualitatively similar dependence beyond $\nu=-2$. The transition at $B_c\approx6$ T for $-2<\nu<-1$ is associated with a change in this background, including a region of enhanced negative $\dudn$ near $\nu=-2$. Negative compressibility also occurs in other localized regions such as at the boundaries separating the slanted stripe of reentrant charge order from the surrounding Hofstadter states (this is more clearly visible at 330 mK, see Supplementary Material). The difference between the background compressibility in regions where Hofstadter states and charge order respectively occur, together with sharper negative $\dudn$ at the boundaries between them, indicate phase transitions and therefore realization of two physically distinct regimes \cite{dultz_thermodynamic_2000,feldman_fractional_2013,zondiner_cascade_2020,zhou_half-_2021,zhou_isospin_2022,yu_correlated_2022-1}.

\section{Interaction-driven moir\'e band crossing}
The interplay between moir\'e band structure and Coulomb interactions is central to understanding the competition between Hofstadter states and charge order. Figure 2A shows a density functional theory calculation of the moir\'e potential landscape in real space. It is characterized by an array of deep potential wells for holes at MX sites interpolated by a smooth potential profile without strong minima at other high-symmetry stackings, realizing a well-defined triangular lattice \cite{shabaniDeepMoirePotentials2021,niekenDirectSTMMeasurements2022}. As a result, electronic states at other sites are disfavored and the lowest few moir\'e bands are well described at the single-particle level by the hybridized eigenstates of an individual potential well (Fig. \ref{fig:band-crossing}B-C). For $s$-orbitals (corresponding to the lowest energy eigenstate of a single well), the hybridization is very weak due to the vanishing inter-site wavefunction overlap, resulting in extremely flat bands with a twofold spin degeneracy. For the four-fold degenerate $p$-orbitals, the larger spatial spread of the wavefunctions leads to stronger overlap and therefore more dispersive bands. At $\nu<-1$, on-site Coulomb repulsion splits the $s$-orbitals into lower and upper Hubbard bands separated by an energy gap $U$. Competition between the single-particle splitting of the noninteracting moir\'e bands and the on-site repulsion controls which band is populated between $-2<\nu<-1$. Experimentally, the magnetic field affects the energetic ordering of the moir\'e bands through spin/orbital Zeeman coupling. The observation of a field-driven phase transition with an abundance of low-field charge ordered states between $-2<\nu<-1$ that are suppressed at high fields provides experimental evidence that the opposite-spin flat ($s$) band is favored at $B=0$, but is close enough to a spin-aligned dispersive ($p$) band such that their energetic ordering can be switched.

The difference in dispersion of the competing bands is a natural seed for the experimentally observed dichotomy of crystalline and fluid ground states, as large effective mass favors Wigner crystals and suppresses cyclotron gaps. We construct a phenomenological Stoner model consisting of flat and dispersive bands, which captures many key features in the experimental data. We simplify the system at fillings $\nu<-1$ as a perfectly flat (upper Hubbard) band and a manifold of dispersive bands with a constant density of states $D$. The free energy of the system can be expressed as:
\begin{equation}
    E(\nu_f, \nu_d, B) = (\Delta - gB)|\nu_d| + \frac{{\nu_d}^2}{2D} + U_{fd}|\nu_f||\nu_d|,
\end{equation}
where $\nu_{d}<0$ and $-1<\nu_f<0$ are the respective fillings of the dispersive and upper flat bands, $\Delta>0$ is the zero-field energy gap between the dispersive band and the upper flat band (Fig. \ref{fig:band-crossing}C), $g$ is the difference of the effective Zeeman coupling constant between the bands, and $U_{fd}$ is a Stoner parameter describing repulsion between the states in the flat and dispersive bands. Minimizing the free energy with respect to $\nu$ and under the constraint $\nu = \nu_f + \nu_d - 1$,  we obtain the band filling phase diagram shown in Fig. \ref{fig:band-crossing}D (see Supplementary Material). 

The charge order and Hofstadter regimes in the experimental data closely coincide with the areas in the phase diagram where flat and dispersive bands are respectively being populated, providing a natural explanation for the distinct behaviors. The model also predicts a region where holes are transferred from the dispersive bands to the flat band (purple, Fig.~\ref{fig:band-crossing}D), which matches the reentrant charge order regime in Fig.~\ref{fig:Fig1}B. Within this region, $|\nu_d|$ decreases as total $|\nu|$ is increased (Fig.~\ref{fig:band-crossing}D, inset). That charge-ordered states occur at the same commensurate fractional fillings in the reentrant regime as for $-2<\nu<-1$ implies simultaneous localization of coexisting light and heavy holes from both flat and dispersive bands into crystalline phases, as opposed to localizing only those from flat bands. The exact nature of these composite charge orders is an interesting open question (see Supplementary Material). While the simplified model does not incorporate microscopic details of the moir\'e bands, it captures the key qualitative features in the data, and demonstrates that energetic competition between flat and dispersive bands gives rise to the experimental phase diagram.

Electronic interaction effects beyond the simplified model modify the details of the phase transition for $-2<\nu<-1$. For example, the abrupt change at $B_c\approx 6$ T is suggestive of a first order transition. Moreover, the gap at $\nu=-2$ does not evolve linearly with magnetic field (Fig. \ref{fig:band-crossing}E). 
Instead, it decreases nonlinearly at low fields, with a sharp drop before stabilizing at high fields. These observations are incompatible with a smooth crossover driven by single-particle energetics.

To understand the sharp transition at $B_c\approx 6$ T in the range $-2<\nu<-1$, we introduce a Hubbard model description of the competing flat and dispersive hole bands. In the limit of a perfectly flat band, this model becomes integrable and can be solved exactly (see Supplementary Material). The resulting phase diagram, computed using the phenomenological parameters used in the Stoner model and in the relevant limit of large Hubbard interaction, is shown in Fig.~\ref{fig:band-crossing}F. The phase diagram indeed shows a sharp transition at $B_c$ associated with sudden depopulation of the flat band. The shape of the phase boundary is in close agreement with the experimental data.  In addition, the region of highly negative compressibility observed near the upturn around $\nu\approx -2$ (Fig.~\ref{fig:Fig1}B) is predicted within the model (see Supplementary Material). Since the ground state at $\nu=-2$ changes abruptly at $B_c$ from a spin-unpolarized to a spin-polarized state, the $\nu=-2$ gap should also abruptly change at $B_c$, as observed experimentally.

The close competition between opposite-spin bands for $-2<\nu<-1$ discussed above requires no fine-tuning and can be understood from general theoretical grounds in the limit of large moir\'e period, for which the field-driven transition is analogous to a singlet-triplet phase transition of a doubly occupied quantum dot. In the absence of interactions, the spin transition occurs when the Zeeman energy gain is sufficient to overcome the gap between moir\'e bands ($\sim 40$ meV from first principles [Supplementary Material]), which would imply an extremely high critical field $B_c$. However, $B_c$ decreases rapidly with interaction strength and falls in an experimentally accessible regime for realistic interaction strengths (see Supplementary Material). Thus, the finite field transition observed at $-2<\nu<-1$ at $B_c\approx 6$ T is a direct consequence of strong electronic interactions. In fact, $B_c$ remains non-zero even in the limit of strong interactions, which follows from the fact that the zero-field ground state of a two-electron system with a spin-independent potential and arbitrarily strong Coulomb interaction is necessarily a spin singlet \cite{lieb_theory_1962}.

Measurements at different spatial locations allow us to address the robustness of our findings and their dependence on twist angle. The experimental phase diagram is qualitatively similar at multiple independent locations with different twist angles (see Supplementary Material). This agreement demonstrates the general applicability of our theoretical framework. Small differences in observed behavior provide further indications regarding the role of twist angle. We find larger twist angle favors an additional Hofstadter state $(-1,-1)$ between $-3/2<\nu<-1$, whereas smaller twist angle suppresses all Hofstadter states for $\nu>-2$. This is consistent with the decreased bandwidth at smaller twist angles and the associated tendency to favor charge order. The critical field $B_c$ also decreases with decreasing $\theta$, consistent with the predictions from the phenomenological and microscopic models (see Supplementary Material). 

\section{Doublon Wigner crystals}
\begin{figure*}[t]
    \renewcommand{\thefigure}{\arabic{figure}}
    \centering
    \includegraphics[scale =1.0]{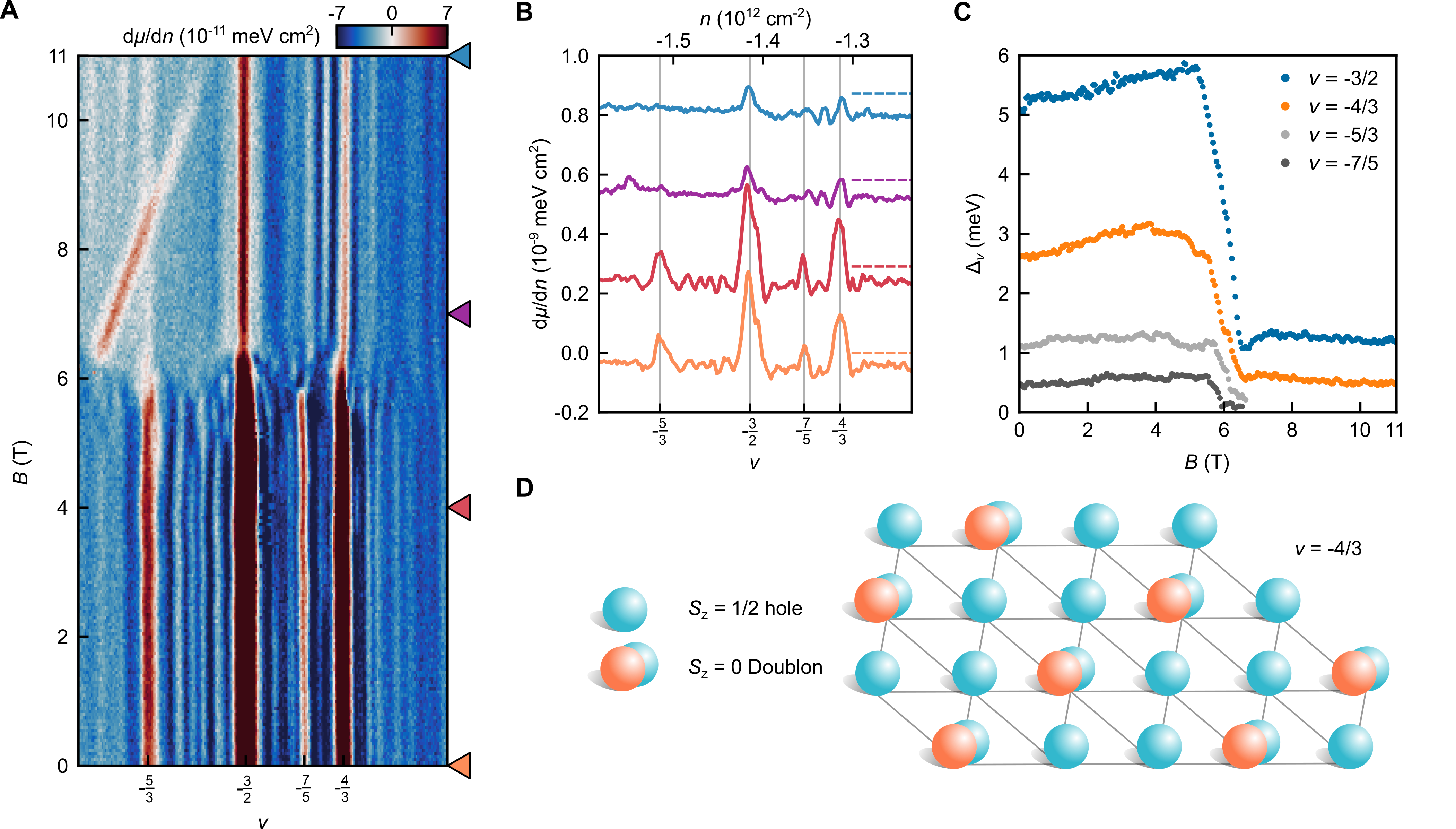}
    \caption{\textbf{Magnetic field dependence of charge-ordered states.} \textbf{(A)} High-resolution measurement of $\dudn$ as a function of $\nu$ and $B$. \textbf{(B)} Line cuts of $\dudn$ at $B=0,4,7$, and 11 T as indicated by the arrows in (A). Each line is offset by $0.28\times10^{-9}$ meV cm$^2$ (dashed lines on the right correspond to $\dudn=0$). \textbf{(C)} Magnetic field dependence of the thermodynamic gaps of charge-ordered states. \textbf{(D)} Illustration of $\nu=-4/3$ charge-ordered state showing a periodic arrangement of doubly occupied sites.}
    \label{fig:ChO-Mag}
\end{figure*}

Having established this general framework, we next present higher-resolution measurements of the charge-ordered states between $-2<\nu<-1$ (Fig. \ref{fig:ChO-Mag}A-C) and discuss them in more detail. While commensurate charge order has been observed in various moir\'e TMD heterobilayers \cite{tangSimulationHubbardModel2020a,reganMottGeneralizedWigner2020,xuCorrelatedInsulatingStates2020a,liChargeorderenhancedCapacitanceSemiconductor2021,jin_stripe_2021,huang_correlated_2021,liu_excitonic_2021,chu_nanoscale_2020,li_imaging_2021}, the unique moir\'e potential profile in MoSe$_2$/WSe$_2$ differentiates the states we observe in essential ways. Crucially, at low magnetic fields, holes occupy the upper Hubbard band at $-2<\nu<-1$, implying that $s$-orbitals at each site of the triangular moir\'e lattice are doubly occupied, forming ‘doublons’ with energy cost $U$. This contrasts with systems in which the second moir\'e band comes from orbitals located in different high-symmetry stacking sites, realizing a charge-transfer insulator without double occupation of the same orbitals \cite{zhangMoirQuantumChemistry2020}. Consequently, the charge-ordered states we observe between $-2<\nu<-1$ necessarily involve periodic arrangements of doublons surrounded by singly occupied sites, realizing a new type of generalized Wigner crystal, which we refer to as a ‘doublon Wigner crystal’ (DWC) (Fig. \ref{fig:ChO-Mag}D). It is likely that the very deep moir\'e potential and large moir\'e unit cell (i.e. flat bandwidth) combine to stabilize the DWCs that we observe \cite{padhi_generalized_2021}. 
 
The thermodynamic gaps $\Delta_{\nu}$ of the DWCs at filling fraction $\nu$ exhibit an unexpected dependence on magnetic field (Fig.~\ref{fig:ChO-Mag}C). The gap at $\nu=-3/2$ is the largest, and surprisingly, we find that it and other DWC gaps grow with increasing $B$ below the critical field $B_c\approx 6$ T. After the transition, the remaining charge-ordered gaps decrease slightly with increasing $B$. Increasing gaps are also observed in other independent locations for certain filling factors, though quantitative details depend on position (see Supplementary Material). This is inconsistent with expectations from a scenario where particle- and hole-like excitations of the DWCs carry the same spin/orbital moments, and implies that the excitations couple differently to the magnetic field. One possibility is that the hole-like excitations involve itinerant doublons dressed by local regions in which an antiferromagnetic spin configuration is favored, thus giving rise to quasiparticle with a different effective spin  \cite{zhang_pairing_2018,davydova_itinerant_2022,zhang2022pseudogap,morera_high-temperature_2022,foutty_tunable_2022}. This is scenario is favored in the large $U/t$ limit of the triangular Hubbard model, which our system realizes \cite{wuHubbardModelPhysics2018,zhangMoirQuantumChemistry2020}. Orbital effects such as a momentum-dependent Berry curvature could induce an orbital moment that could also give rise to different effective Zeeman coupling between particle- and hole-like excitations \cite{xiao_valley-contrasting_2007}. Our results motivate additional experimental and theoretical work to distinguish these possibilities. 

\section{Hofstadter patchwork at higher hole doping}

\begin{figure*}[t]
    \renewcommand{\thefigure}{\arabic{figure}}
    \centering
    \includegraphics[scale =1.0]{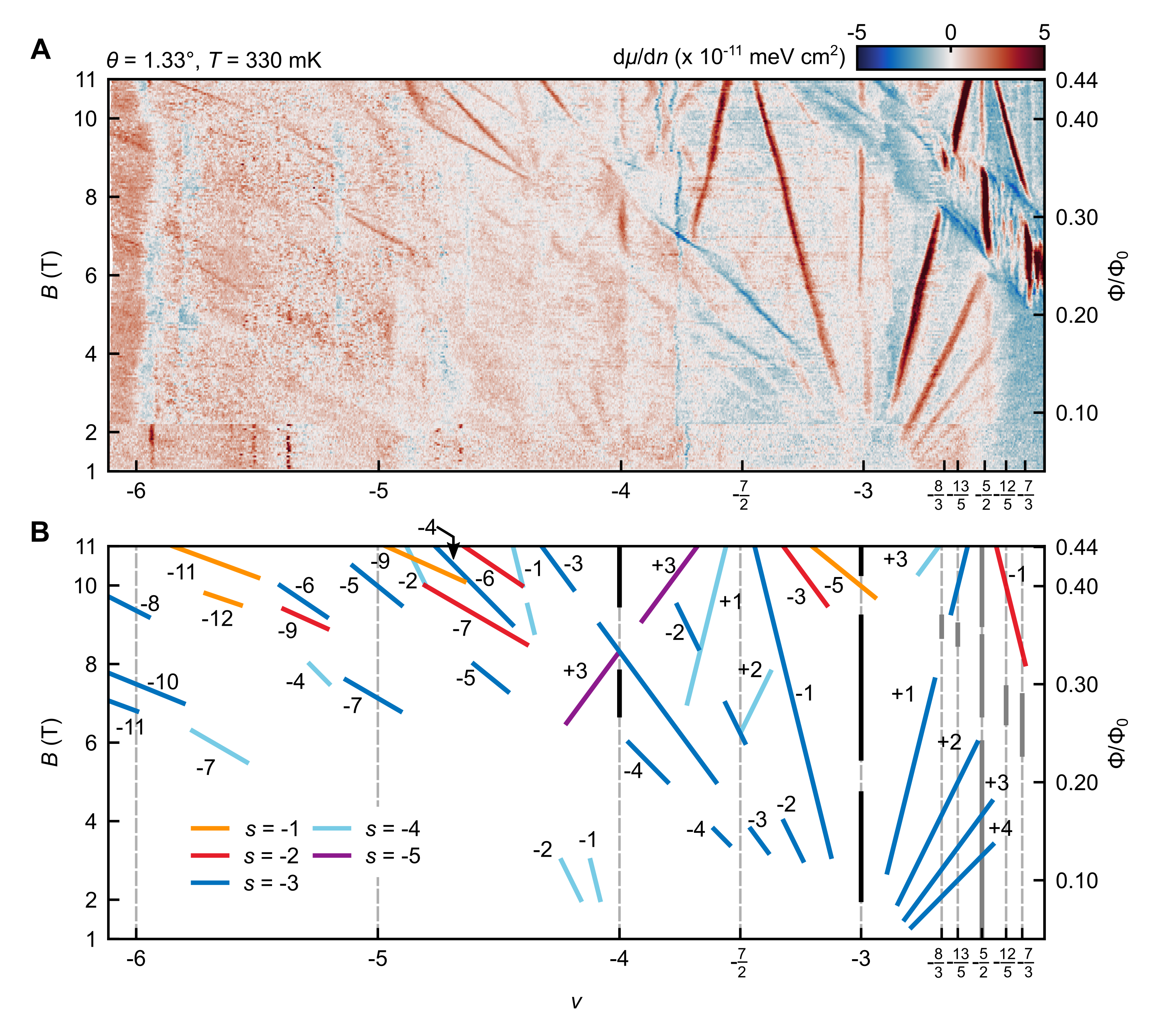}
    \caption{\textbf{Interpenetrating Hofstadter states.} \textbf{(A)} Higher resolution measurement of $\dudn$ as a function of $B$ over a larger range of $\nu$ at $T=330$ mK. \textbf{(B)} Incompressible states identified from (A). Chern numbers $t$ of Hofstadter states are labeled, and the $s$ for each state is indicated by their line colors (legend).} 
    \label{fig:high-dope}
\end{figure*}

The large moir\'e unit cell of MoSe$_2$/WSe$_2$ renders hole doping beyond $\nu=-6$ accessible, allowing us to probe higher moir\'e bands in the Hofstadter regime. Figure \ref{fig:high-dope}A shows $\dudn$ in a large hole doping range $\nu<-2$ at $T=330$ mK, measured at a location near that of Fig. \ref{fig:Fig1}B with slightly decreased tip height above the sample. The reduced thermal and spatial broadening reveals a significantly richer pattern of Hofstadter and charge-ordered states whose $(t,s)$ are identified in Fig. \ref{fig:high-dope}B. At high fields, additional incompressible states form sequences emanating downward from commensurate flux $\Phi/\Phi_0=1/2$ and $\nu=-5/2, -7/2$, clearly indicating a Hofstadter origin. The most prominent features at low fields remain the singly-degenerate Landau fan surrounding $\nu=-3$, with weaker incompressible states having negative $t$ also visible. 

We observe multiple phase transitions that result from energetic reordering of Hofstadter subbands, leading to changes in their occupations. States within the negative-$t$ Landau fan from $\nu=-3$ terminate at a series of negative compressibility features at intermediate fields, indicating a cascade of phase transitions. Similarly, gaps at $\nu = -3/2, -3$, and $-4$ all exhibit transitions in which they close and reopen as a function of magnetic field. Competition between different subband occupancies is especially prominent at higher doping $\nu<-4$, where multiple interpenetrating Hofstadter states with $s=-1, -2, -3, -4$ and $-5$ appear over narrow ranges in density and magnetic field. Remarkably, the filling at which some of these Hofstadter states occur deviates from their $B=0$ intercept by more than five filling factors. This pattern is in sharp contrast to the flavor symmetry broken Hofstadter-Chern insulators observed in magic-angle twisted bilayer graphene systems \cite{saito_hofstadter_2021,park_flavour_2021,nuckolls_strongly_2020,choi_correlation-driven_2021,yu_correlated_2022-1}, which are generally bounded between $|s|$ and $|s+1|$. It does, however, resemble the Hofstadter pattern observed in hBN-aligned graphene devices \cite{ponomarenko_cloning_2013,dean_hofstadters_2013,hunt_massive_2013,wang_evidence_2015,spantonObservationFractionalChern2018}, where such a bound appears absent.

Certain aspects of the Hofstadter spectrum match single-particle expectations. For example, the Laudau fan asymmetry between $-3<\nu<-2$ is qualitatively consistent with single-particle Hofstadter calculations, which show a pronounced asymmetry of effective masses for opposite band edges of the first dispersive band (see Supplementary Material). The single-particle calculations are also consistent with the absence of gaps at integer fillings $\nu \le -4$ at low magnetic fields, where Hofstadter subbands of the dispersive moir\'e bands overlap. However, other observations clearly point to interaction effects. The patchwork of Hofstadter states with different $s$ at high density reflects close competition of single-particle and interaction effects. Namely, the narrow bandwidth and small minigaps in the Hofstadter spectrum render the system susceptible to filling-dependent interaction energies. This causes the ordering of Hofstadter subbands originating from different moir\'e bands to reshuffle in a highly non-monotonic fashion. 


\section{Conclusion}
In conclusion, we have demonstrated coexisting Hofstadter and charge-ordered states in a twisted MoSe$_2$/WSe$_2$ heterobilayer, bridging two distinct physical regimes in a single system. Our measurements highlight the relevance of multiple moir\'e bands of differing character whose energetic ordering can be tuned by laboratory magnetic fields. The coexistence of dispersive and flat bands may provide a venue to investigate Kondo lattice physics \cite{zhao_gate-tunable_2022}. The large moir\'e unit cell introduces additional opportunities, as long-range and even on-site Coulomb interactions can be potentially controlled by placing a nearby screening layer \cite{liu_tuning_2021}. This may enable the study of melting of (generalized) Wigner crystals \cite{matty_melting_2021}, as well as provide a method to realize more exotic quantum phases of matter, including fractional Chern insulators \cite{paul_moire_2022,crepel_anomalous_2022}, Nagaoka ferromagnets  \cite{arovas_hubbard_2022}, and quantum spin liquids \cite{arovas_hubbard_2022}. 

\section{References and Notes}
\vspace{-0.4in}
\bibliographystyle{apsrev4-1}
\bibliography{references.bib, references-2.bib}

\section{Acknowledgements}
We thank Tony Heinz, Aidan O'Beirne, and Henrique B. Ribeiro for their assistance with second harmonic generation measurements, and Thomas P. Devereaux and Alexander A. Zibrov for helpful discussions.

\section{Funding}
Experimental work was primarily supported by NSF-DMR-2103910. B.E.F. acknowledges an Alfred P. Sloan Foundation Fellowship and a Cottrell Scholar Award. The work at MIT was funded by the Air Force Office of Scientific
Research (AFOSR) under award FA9550-22-1-0432. K.W. and T.T. acknowledge support from JSPS KAKENHI (Grant Numbers 19H05790, 20H00354 and 21H05233). B.A.F. acknowledges a Stanford Graduate Fellowship. Part of this work was performed at the Stanford Nano Shared Facilities (SNSF), supported by the National Science Foundation under award ECCS-2026822.

\section{Author contributions}
C.R.K, J.Y. conducted the scanning SET measurements. C.R.K, J.Y., and B.E.F. designed the experiment. T.D., A.P.R., Y.Z. and L.F. conducted theoretical calculations. C.R.K. fabricated the sample. K.W. and T.T. provided the hBN crystals. B.E.F. and L.F. supervised the project. All authors participated in analysis of the data and writing of the manuscript. 

\section{Competing interests}
The authors declare no competing interest. 

\section{Data and materials availability}
The data that supports the findings of this study are available from the corresponding authors upon reasonable request.

\end{document}


\title{Supplementary Information for: \\Hofstadter states and reentrant charge order in a semiconductor moir\'e lattice}

\author{Carlos R. Kometter}
\thanks{These authors contributed equally}
\affiliation{Geballe Laboratory for Advanced Materials, Stanford, CA 94305, USA}
\affiliation{Department of Physics, Stanford University, Stanford, CA 94305, USA}

\author{Jiachen Yu}
\thanks{These authors contributed equally}
\affiliation{Geballe Laboratory for Advanced Materials, Stanford, CA 94305, USA}
\affiliation{Department of Applied Physics, Stanford University, Stanford, CA 94305, USA}

\author{Trithep Devakul}
\thanks{These authors contributed equally}
\affiliation{Department of Physics, Massachusetts Institute of Technology, Cambridge, Massachusetts 02139, USA}

\author{Aidan P. Reddy}
\affiliation{Department of Physics, Massachusetts Institute of Technology, Cambridge, Massachusetts 02139, USA}

\author{Yang Zhang}
\affiliation{Department of Physics, Massachusetts Institute of Technology, Cambridge, Massachusetts 02139, USA}

\author{Benjamin A. Foutty}
\affiliation{Geballe Laboratory for Advanced Materials, Stanford, CA 94305, USA}
\affiliation{Department of Physics, Stanford University, Stanford, CA 94305, USA}

\author{Kenji Watanabe}
\affiliation{Research Center for Functional Materials, National Institute for Material Science, 1-1 Namiki, Tsukuba 305-0044, Japan}

\author{Takashi Taniguchi}
\affiliation{International Center for Materials Nanoarchitectonics, National Institute for Material Science, 1-1 Namiki, Tsukuba 305-0044, Japan}

\author{Liang Fu}
\affiliation{Department of Physics, Massachusetts Institute of Technology, Cambridge, Massachusetts 02139, USA}

\author{Benjamin E. Feldman}
\email{bef@stanford.edu}
\affiliation{Geballe Laboratory for Advanced Materials, Stanford, CA 94305, USA}
\affiliation{Department of Physics, Stanford University, Stanford, CA 94305, USA}
\affiliation{Stanford Institute for Materials and Energy Sciences, SLAC National Accelerator Laboratory, Menlo Park, CA 94025, USA}

\maketitle

\begin{center}
\vspace{1 in}
This PDF file includes:

Materials and Methods\\
Supplementary Text\\
Supplementary Figs. S1 to S13
\end{center}

\newpage

\tableofcontents

\supplementarysection
\section{Materials and Methods}

\subsection{Sample fabrication}
\begin{figure}[b!]
    \centering
    \includegraphics[scale =1.0]{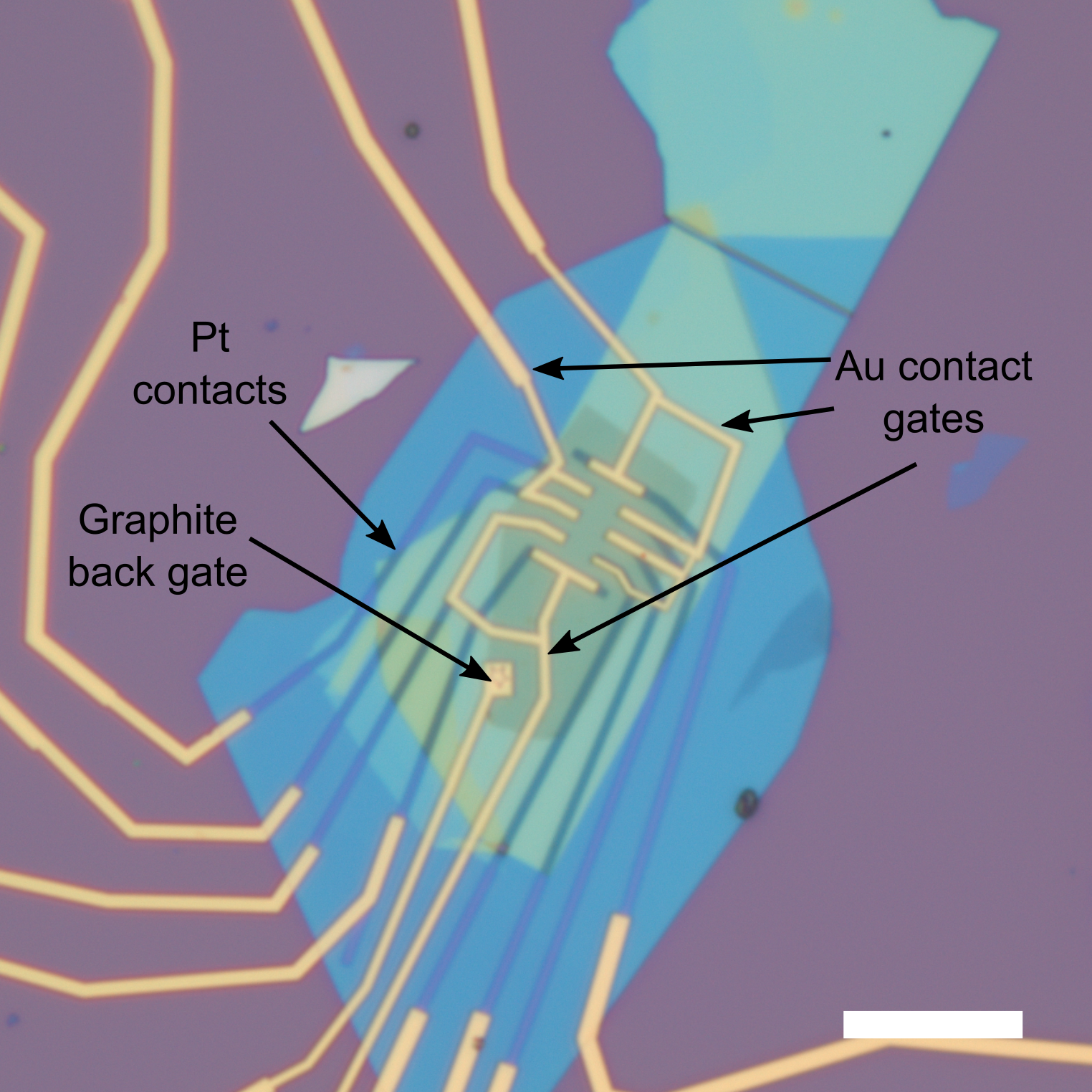}
    \caption{\textbf{Device image.} Optical microscope image of the MoSe$_2$/WSe$_2$ device. Platinum contacts to the sample, gold `contact' gates to locally dope the heterobilayer, and the back gate electrode are all indicated. Scale bar: 10 $\mu$m.} 
    \label{fig:device}
\end{figure}
The twisted MoSe$_2$/WSe$_2$ device (Fig. \ref{fig:device}) was fabricated using standard dry transfer techniques \cite{wang_one-dimensional_2013}. The MoSe$_2$ and WSe$_2$ monolayers, few-layer graphite, and hexagonal boron nitride (hBN) flakes were exfoliated from bulk crystals. All pickup steps were performed with poly(bisphenol A carbonate) (PC) / polydimethylsiloxane (PDMS) stamps. First, pre-patterned Pt contacts (5 nm thickness) were evaporated using standard electron beam (e-beam) lithography and metallization techniques onto a 30-nm hBN previously stacked on few-layer graphite. Annealing was performed at 400$\degree$C for 8 hours before and after pre-patterning to eliminate any polymer residue. A 30-nm hBN flake was then used to pick up monolayers of MoSe$_2$ and WSe$_2$, which were deposited atop the pre-patterned contacts to complete the stack. Finally, e-beam lithography, CHF$_3$/O$_2$ etching, and metal deposition were used to pattern local `contact' gates above the Pt contact area, and to make electrical contact to the Pt leads and graphite gate with Cu/Au (2~nm/100~nm). The `contact' gates served to locally dope the transition metal dichalcogenide (TMD) to achieve Ohmic contact \cite{movvaHighMobilityHolesDualGated2015}.

\subsection{Single-electron transistor (SET) fabrication and measurement}

The single-electron transistor (SET) sensor was fabricated by evaporating aluminium onto the apex of a pulled quartz rod. The size of the apex, and thus the lateral dimension of the SET, is estimated to be 50-100 nm based on scanning electron microscope imaging of tips fabricated using the same procedure. It was brought approximately 50 nm above the sample surface, resulting in an overall spatial resolution of about 100-150 nm. The scanning SET measurements were performed in a Unisoku USM 1300 scanning probe microscope with a customized microscope head. An AC excitation of $V_\textrm{G,AC}=4-8$ mV at frequency $f_\textrm{G}=233.33$ Hz was applied to the back gate, and an AC excitation $V_\textrm{2D,AC}=5$ mV at frequency $f_\textrm{2D}=167.77$ Hz was applied to the sample. We then measured the inverse compressibility $\dudn \propto \frac{I_\textrm{G}/V_\textrm{G,AC}}{I_\textrm{2D}/V_\textrm{2D,AC}}$, where $I_\textrm{G}$ and $I_\textrm{2D}$ are demodulated from the SET current through the SET probe using standard lock-in techniques. A DC offset voltage $V_\textrm{2D,DC}$ is further applied to the sample to maintain the SET at its maximum sensitivity point within a Coulomb blockade oscillation fringe chosen to be near the `flat-band' condition where the tip does not gate the sample. This minimizes tip-induced doping and provides an independent direct DC measurement of $\mu(n)$. If the sample or contacts are highly resistive such that the inverse measurement frequency is short compared to the RC charging time of the sample, the AC measurement of inverse compressibility can be artificially enhanced. The DC measurement is noisier but less susceptible to these effects. For the data presented in this manuscript, the AC and DC measurements agree within the noise, indicating sufficient charging even in the AC scheme. We therefore obtain $\mu(n)$ by integrating the lower-noise AC measurement of $\dudn$. To reduce contact resistance and facilitate sample charging, a large negative voltage $V_\textrm{CG}$ is applied to the `contact' gates. All measurements were performed at temperature $T=1.6$ K unless specified otherwise.

\subsection{Determination of stacking configuration}
\begin{figure}
    \centering
    \includegraphics[scale =1.0]{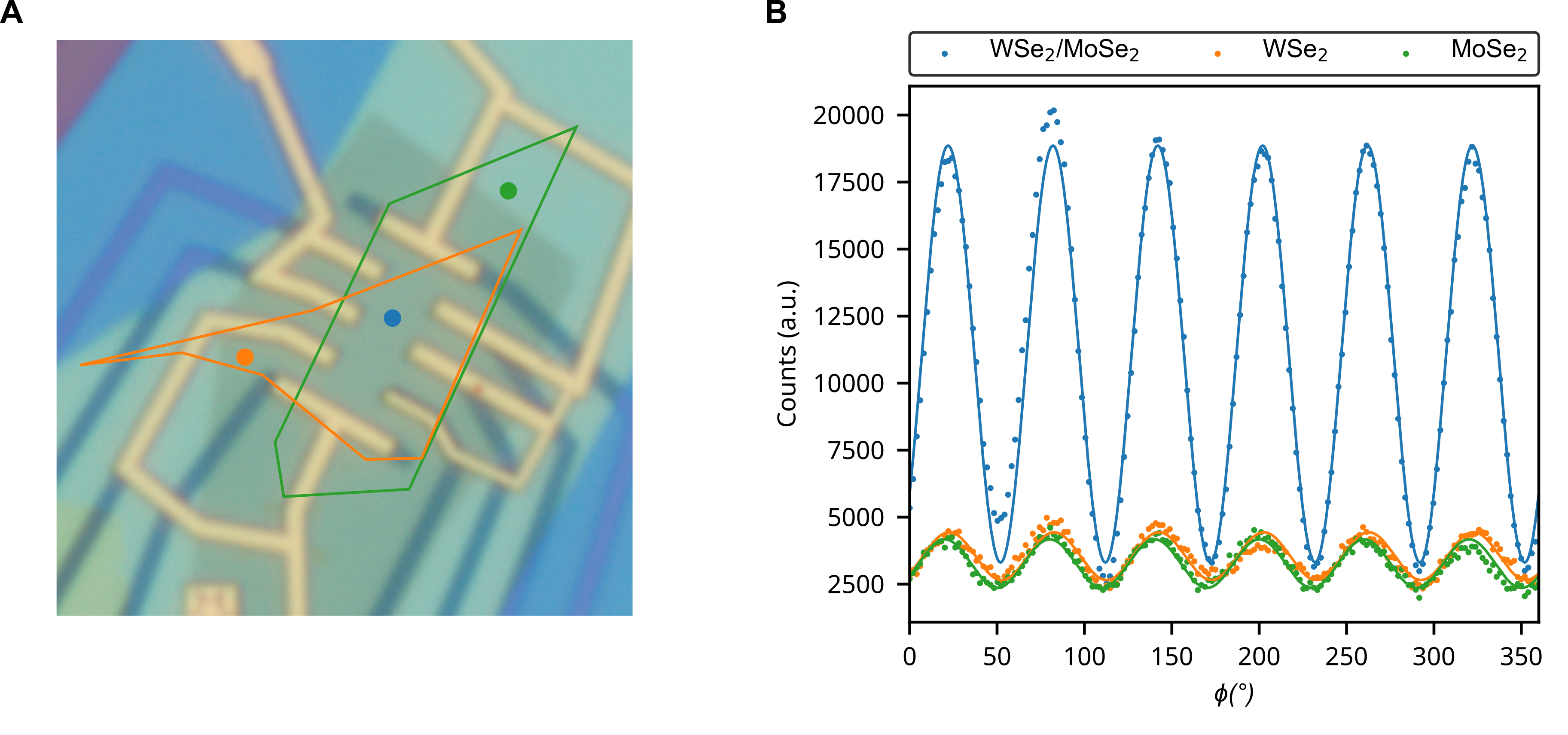}
    \caption{\textbf{Second harmonic generation (SHG) measurements.} \textbf{(A)} Zoom-in of the device with WSe$_2$ (orange) and MoSe$_2$ (green) monolayers outlined. The colored circles indicate the locations where SHG measurements were taken. \textbf{(B)} SHG signal from each isolated monolayer and the heterobilayer region as a function of polarization angle $\phi$. The larger signal from the heterobilayer indicates that the constituent layers have AA stacking.} 
    \label{fig:shg}
\end{figure}
Second harmonic generation (SHG) can be used to determine the relative twist between monolayer WSe$_2$ and monolayer MoSe$_2$ due to their lack of inversion symmetry \cite{kumarSecondHarmonicMicroscopy2013,liProbingSymmetryProperties2013}. By tracking the reflected signal at the second harmonic frequency of the incoming signal as the linear polarizer rotates, crystallographic orientations of monolayer portions of the device are identified as shown in Fig.~\ref{fig:shg}. The SHG signal also clarifies the stacking of the heterobilayer. For a bilayer, the second harmonic fields add constructively for AA stacking and destructively for AB stacking. The strong SHG signal over the heterobilayer portion of the device demonstrates that it has AA stacking configuration (Fig.~\ref{fig:shg}B). 

\subsection{Determination of the twist angle and filling factors}
For homobilayers and heterobilayers with small lattice mismatch, the moir\'e unit cell area diverges as the twist angle approaches zero. Twist angle disorder can therefore generate local variability in the unit cell area. Thus, while SHG measurements distinguish between AA and AB stacking, we apply a more precise method to estimate the local twist angle in SET measurements. We calibrate the gate-sample capacitance and therefore the conversion between back gate voltage $V_\textrm{G}$ and carrier density $n$ based on the slope of the Hofstadter states in a magnetic field $B$. We then identify one and two holes per moir\'e unit cell ($\nu=-1,-2$) as the densities with the largest peaks in inverse compressibility (the relative gate voltages they occur at in different locations with distinct twist angles further confirms this filling factor assignment). The $B=0$ intercept obtained from extrapolation of the Hofstadter states also matches these assignments and confirms that the largest peaks are at integer fillings. From the density at $\nu=-2$, we calculate the twist angle according to  $n=\frac{4}{\sqrt{3}}\frac{\theta^2+\delta^2}{a_{\text{Mo}}^2}$, where $\delta=1-\frac{a_\text{W}}{a_\text{Mo}}$ and $a_{\text{W(Mo)}}= 3.282\text{ \AA}\ (3.290 \text{ \AA})$ is the lattice constant for WSe$_2$ (MoSe$_2$).

\subsection{Gap extraction}

\begin{figure*}[t]
    \centering
    \includegraphics[scale =1.0]{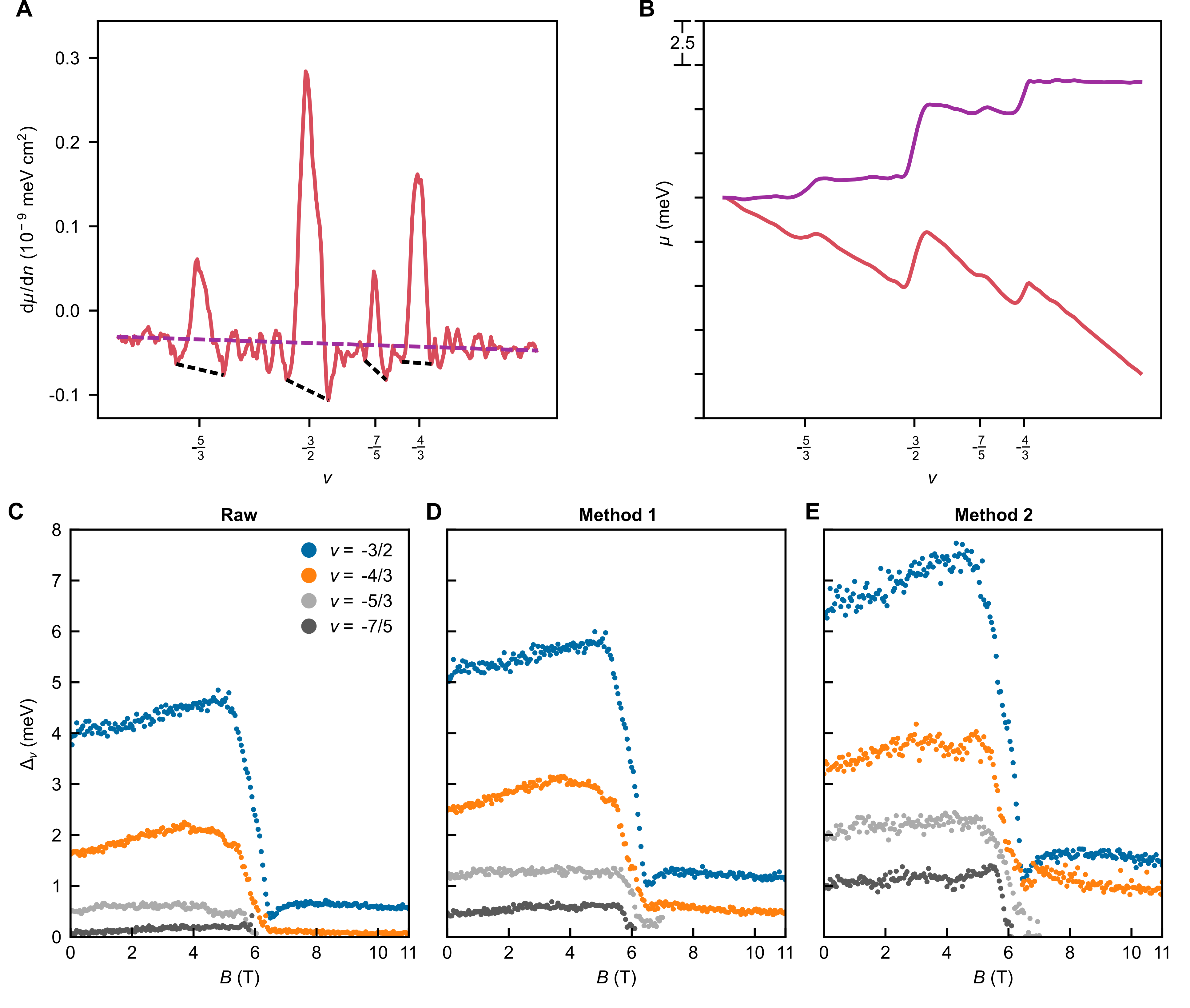}
    \caption{\textbf{Inverse electronic compressibility $\dudn$ background and determination of gaps at fractional fillings.} \textbf{(A)} Local $\dudn$ as a function of moir\'e filling factor $\nu$ (red). Purple and black dashed lines correspond to Methods 1 and 2, respectively, of accounting for the negative compressibility background. The purple dashed line is a linear fit to the overall negative background outside the peaks (Method 1), while the black dashed lines linearly connect the local minima surrounding the incompressible peaks of the charge-ordered states (Method 2). \textbf{(B)} Chemical potential $\mu$ as a function of $\nu$ from the raw $\dudn$ (red) and after subtracting the purple linear fit from (A) (purple). (\textbf{C}-\textbf{E}) The gap size $\Delta_{\nu}$ of each charge-ordered state as a function of magnetic field $B$. These correspond to the step size in chemical potential at each respective $\nu$ from the raw $\dudn$ (C), that obtained using Method 1 to account for the negative background (D), and that determined from Method 2 (E).} 
    \label{fig:neg_dudn_rm}
\end{figure*}

The thermodynamic gap $\Delta_{\nu}$ of a given incompressible state at filling $\nu$ is given by the corresponding step in chemical potential. We extract the energy gaps by integrating $\dudn$ with respect to the density $n$,
\begin{align*}
    \Delta_{\nu}\equiv\Delta\mu&=\mu(n_+)-\mu(n_-)\\
    &=\int^{n_+}_{n_-}\left[\frac{\textrm{d}\mu}{\textrm{d} n}-\left(\frac{\textrm{d}\mu}{\textrm{d} n}\right)_{B}\right]  \text{d}n,
\end{align*}
where $n_{+(-)}$ is the density at the local maximum (minimum) chemical potential above (below) the state of interest, and $\left(\frac{\textrm{d}\mu}{\textrm{d}n}\right)_{B}$ is the negative compressibility background to be removed.

As noted in the main text, there is a strong background of negative compressibility over large ranges of filling factor and magnetic field. Simply integrating the raw measured $\dudn$ would therefore result in gaps that are suppressed relative to their true value. To account for the background and obtain more accurate gap sizes, we implement a method (Method 1) analogous to that followed in Ref.~\cite{eisensteinCompressibilityTwodimensionalElectron1994}. Specifically, a linear fit to the $\dudn$ background surrounding the incompressible charge ordered states at fractional filling is subtracted from the raw $\dudn$ (purple dashed line, Fig.~\ref{fig:neg_dudn_rm}A). This method only assumes a slowly varying negative background and is insensitive to details in the vicinity of the incompressible peaks. The chemical potential before and after accounting for the negative background is shown in red and purple, respectively, in Fig.~\ref{fig:neg_dudn_rm}B. The corresponding gaps extracted from the raw and background-corrected chemical potential are plotted as a function of magnetic field in Fig.~\ref{fig:neg_dudn_rm}C and Fig.~\ref{fig:neg_dudn_rm}D, respectively.

Finally, for completeness we also show a different method (Method 2) of accounting for the background. Rather than linearly fitting to the background far from the peaks, we identify the local minima in $\dudn$ adjacent to each incompressible peak (black dashed lines, Fig.~\ref{fig:neg_dudn_rm}A). Integrating the area under the peak but above the line connecting these local minima yields the energy gaps shown in Fig.~\ref{fig:neg_dudn_rm}E. The gaps extracted from the raw $\dudn$ provide an effective lower bound, and those extracted using Method 2 serve as an upper bound. We believe that Method 1 is the most physically justified, and present the corresponding gaps in Fig. 3 of the main text. We emphasize that the method used to account for the negative background compressibility changes the gap sizes quantitatively but not qualitatively. Specifically, for all three gap extraction methods, the gaps of charged ordered states at fractional fillings increase with $B$ at magnetic fields below the phase transition at 6 T and decrease with $B$ above this critical field. 



\section{Comparison between transport and compressibility measurements}
\begin{figure}[h!]
    \centering
    \includegraphics[scale =1.0]{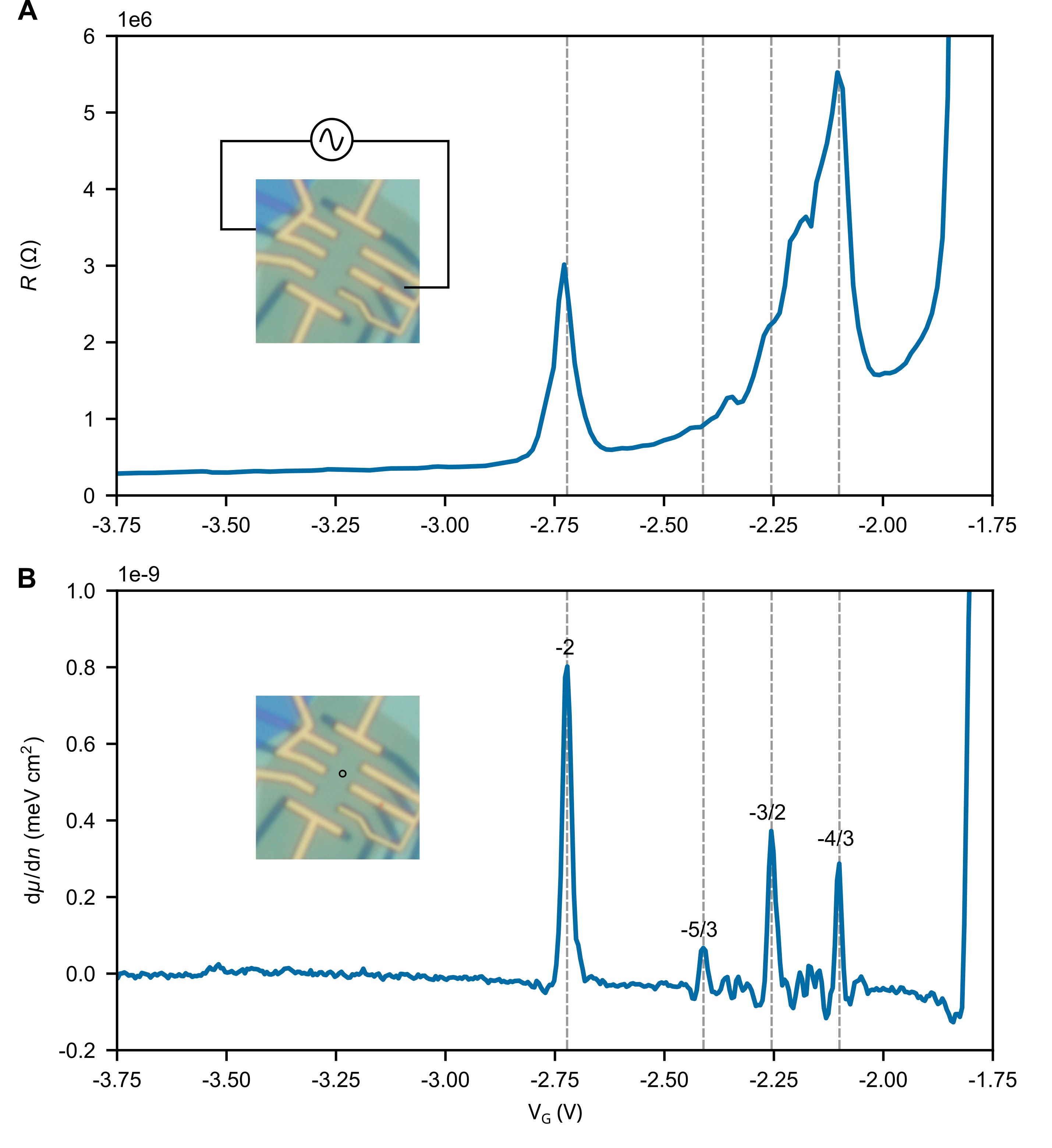}
    \caption{\textbf{Comparison of transport and $\dudn$.} \textbf{(A)} Two-probe resistance $R$ as a function of back gate voltage $V_{\textrm{G}}$ at temperature $T=4$ K. Inset: Circuit diagram showing the contacts used for the measurement. \textbf{(B)} Local $\dudn$ measurement at a location with twist angle $\theta=1.35\degree$ at temperature $T=1.6$ K. Inset: The approximate tip location of this measurement is indicated by the circle.} 
    \label{fig:twoprobe}
\end{figure}

Transport measurements across the device at $B = 0$ (Fig.~\ref{fig:twoprobe}A) show prominent resistance peaks that can be matched to gapped states visible in SET measurements conducted in an intervening region of the sample (Fig.~\ref{fig:twoprobe}B). However, the local electronic compressibility measurements are significantly sharper and reveal additional charge ordered states at fractional fillings. The mobility edge in transport also occurs further from the band edge than in SET measurements. 

\section{Spatial dependence}
\begin{figure}[h!]
    \centering
    \includegraphics[scale =1.0]{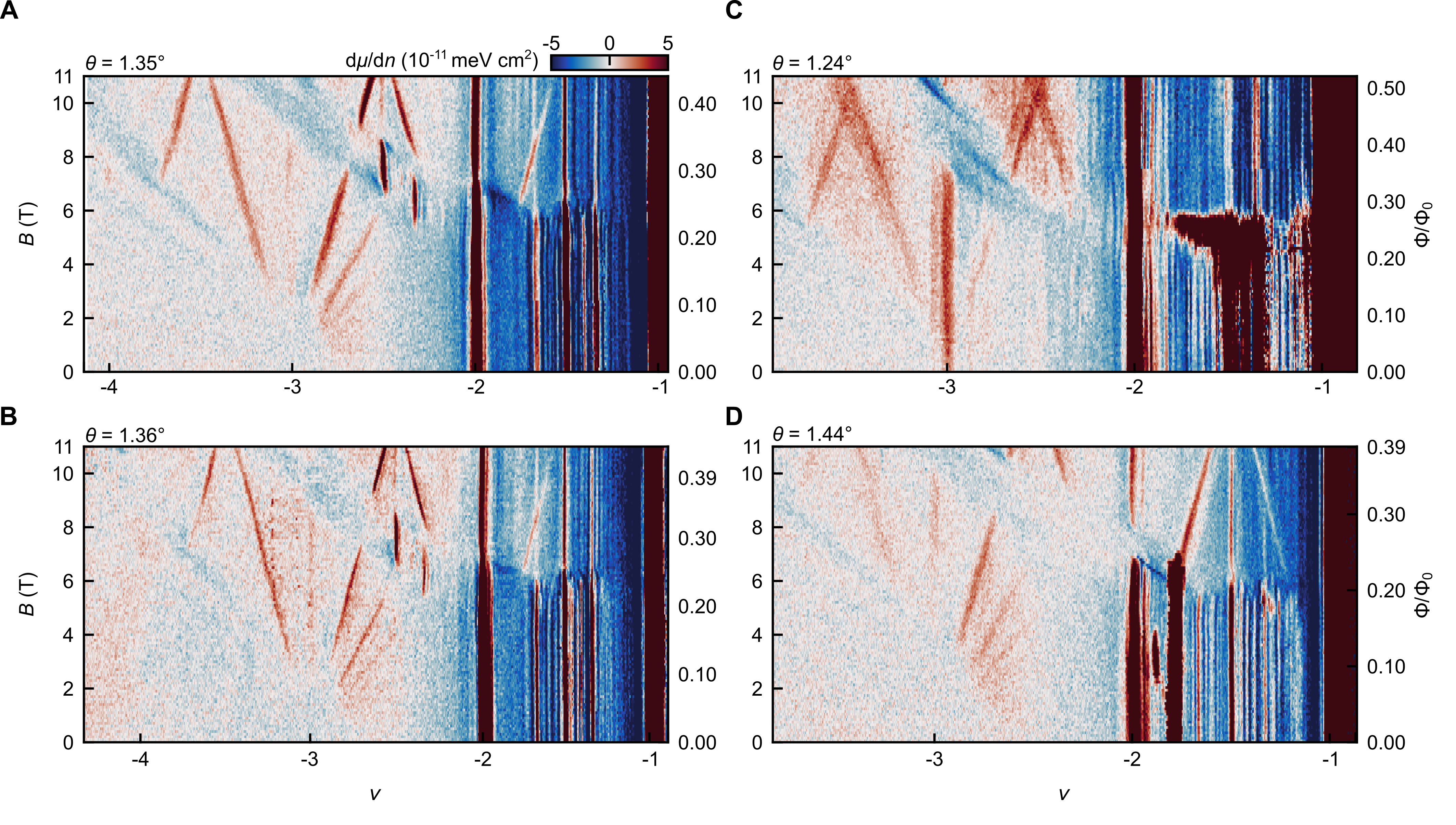}
    \caption{\textbf{Spatial dependence.} (\textbf{A}-\textbf{D}) Measurements of local $\dudn$ as a function of $\nu$ and $B$ at four independent locations which are each separated by at least 150 nm from each other and from the data presented in Fig.~1B of the main text. The corresponding local twist angles are $1.35\degree$ (A), $1.36\degree$ (B), $1.24\degree$ (C), and $1.44\degree$ (D). A qualitatively similar phase diagram of competing Hofstadter and charged-ordered states is evident in each location, demonstrating the generality of the underlying physics. In (D), a constant background compressibility of $8\times10^{-12}$ meV cm$^2$ was subtracted from the raw AC signal, which exhibited a spurious enhancement due to poorer sample charging. The subtracted background was benchmarked based on a simultaneous DC measurement, and the uncertainty of this background is $<2\times10^{-12}$ meV cm$^2$.} 
    \label{fig:Fig2}
\end{figure}
The general pattern of Hofstadter and charge-ordered phases are not unique to a particular twist angle or location in our device. Measurements of inverse compressibility as a function of carrier density and magnetic field at five different locations of the sample demonstrate the robustness of these phases and their phenomenology (Fig. 1b and Fig.~\ref{fig:Fig2}). All of these locations show magnetic field driven phase transitions indicated by an abrupt decrease of the $\nu=-2$ gap and weakened or absent charge-ordered states between $-2<\nu<-1$ at high field, with near perfect agreement in two independent locations that have almost identical twist angles (Fig.~\ref{fig:Fig2}A,B). Other features, such as regions of negative compressibility and the overarching pattern of where Hofstadter and charge-ordered states respectively dominate, are linked to filling factor and occur at similar locations within the phase diagram, showing that they are caused by intrinsic local physics in the heterobilayer and do not simply occur at a specific gate voltage or field. 

Beyond this general agreement, we note qualitative differences at larger and smaller twist angles (Fig.~\ref{fig:Fig2}C,D). At smaller twist angles, we resolve fewer Hofstadter states emanating from $s=-3$, and the $t=1$, $s=-2$ state is absent. In addition, the $\nu=-3$ gap is stronger, appears for a larger range in $B$, and persists all the way down to $B=0$ (Fig.~\ref{fig:Fig2}C). In contrast, at larger twist angles, we observe an extra $t=-1$, $s=-1$ state (Fig.~\ref{fig:Fig2}D) and a $\nu=-2$ gap that closes fully at intermediate fields. Thus, the twist angle tunes the competition between Hofstadter and charge-ordered states. We ascribe this to differences in the electronic bandwidth, which is highly sensitive to the twist angle. Smaller twist angles produce flatter bands, favoring charge-ordered phases, while the more dispersive bands at larger twist angles favor Hofstadter states. 

\begin{figure}[h!]
    \centering
    \includegraphics{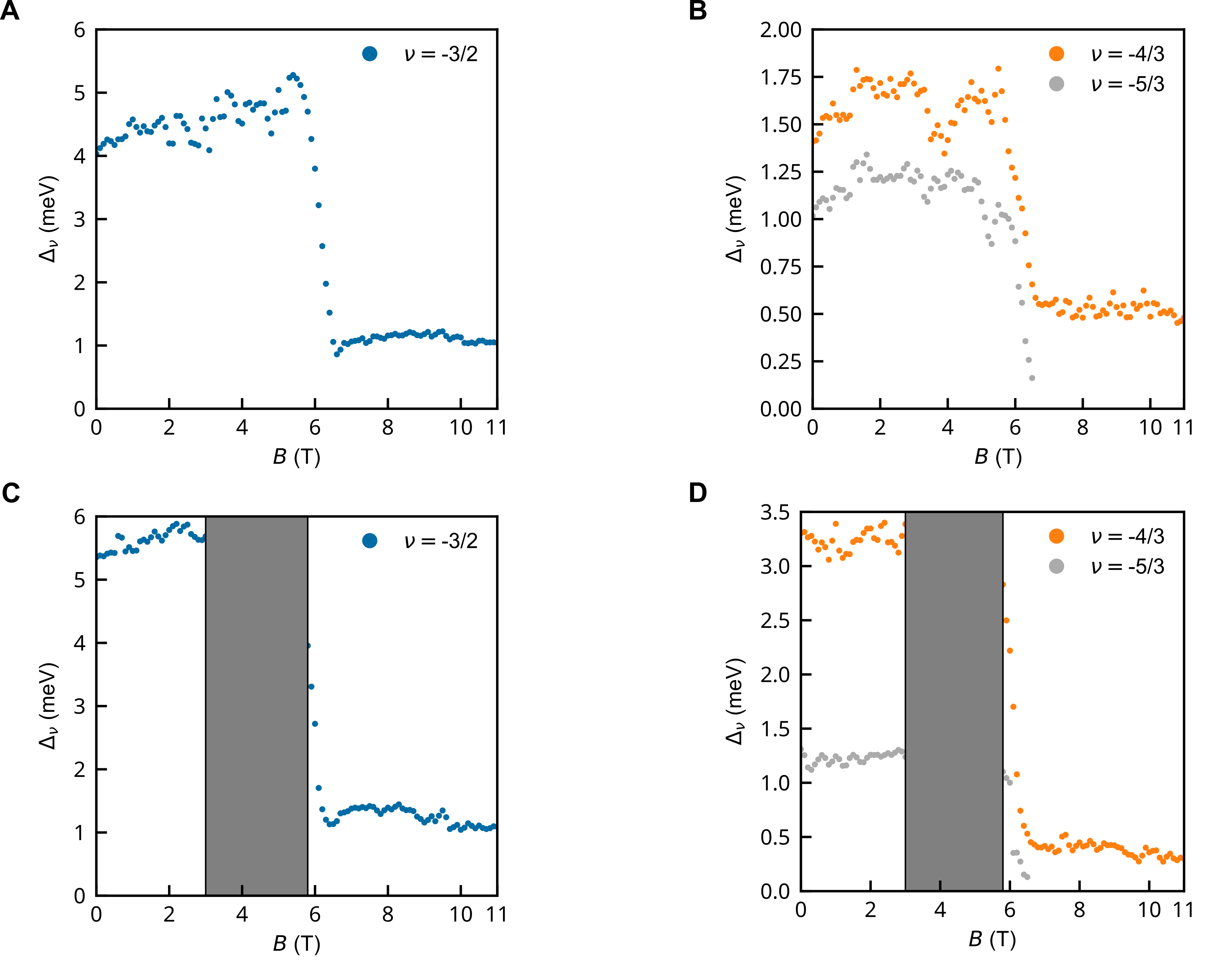}
    \caption{\textbf{Magnetic field dependence of charge-ordered states at other locations.} \textbf{(A,B)} Gaps at fractional fillings extracted from Fig.~1B in the main text. \textbf{(C,D)} Gaps at fractional fillings extracted from Fig.~\ref{fig:Fig2}B. Gray regions indicate fields at which we cannot extract quantitatively accurate gaps because the lock-in amplifier overloaded during measurement.}
    \label{fig:gaps-cho-loc1}
\end{figure}

In Fig.~\ref{fig:gaps-cho-loc1}, we show the magnetic field dependence of the thermodynamic gaps of charge-ordered states in two different locations (those shown in Fig.~1B of the main text and in Fig.~\ref{fig:Fig2}B) that have similar twist angle to Fig.~3C but are sufficiently far away to represent independent measurements. Contradictory to theoretical predictions for a single band Hubbard model in a triangular lattice \cite{panQuantumPhaseDiagram2020,morales-duranMagnetismQuantumMelting2022}, the $\nu=-3/2$ gap is generally larger than those at $-4/3$ and $-5/3$ in our device. This contrasts with several previous measurements in other materials systems where states at multiples of $\nu = 1/3$ are strongest \cite{tangSimulationHubbardModel2020a,liChargeorderenhancedCapacitanceSemiconductor2021,jin_stripe_2021,liContinuousMottTransition2021}. However, we note that a larger gap at $\nu=-3/2$ has also been previously reported in measurements of aligned WS$_2$/WSe$_2$ heterobilayers between $1<|\nu|<2$ \cite{xuCorrelatedInsulatingStates2020a,huangCorrelatedInsulatingStates2021}.

Details of the magnetic field evolution of charge-ordered states at fractional fillings depend on exact location. The increase in the charge gap at $\nu=-3/2$ as a function of magnetic field is reproducible in multiple locations. However, it is harder to identify generic behavior for other fillings. In particular, the gaps at $-4/3$ and $-5/3$ in Fig.~\ref{fig:gaps-cho-loc1}B increase up to $B=1.6$~T beyond which the gap dependence is more ambiguous. In Fig.~\ref{fig:gaps-cho-loc1}D, there is not a large change in gap size over the magnetic fields where we can extract reliable data. We note that while spin physics has been predicted for ground states on a triangular moir\'e lattice \cite{wuHubbardModelPhysics2018,panQuantumPhaseDiagram2020,morales-duranMagnetismQuantumMelting2022}, the very large $U/t \approx O(10^2)$ implies that the energy scale associated with the spin physics, $J=t^2/U \approx 1$ mK, is much smaller than the experimental temperature, and is therefore unlikely to be relevant to the gap scaling with magnetic field. An applied magnetic field squeezes the wave function of an electron in a harmonic potential well, which decreases $|t|$ by reducing the inter-well wavefunction hybridization. This effect could in principle lead to an increase in the gaps of the charge-ordered states with field. However, for realistic parameters of this system, the cyclotron frequency $\omega_c = eB/(2\pi m^*) \ll \omega_0$ is very small ($\omega_0$ is the natural frequency of the harmonic potential), indicating that this effect is also negligible.  

\section{Density function theory (DFT) and Hartree-Fock (HF) calculations}
We study the TMD heterobilayer WSe$_2$/MoSe$_2$ with a small twist angle starting from AA stacking, where every metal (W) or chalcogen (Se) atom  on  the top layer is aligned with the same type of atom on the bottom layer. Within a local region of a twisted bilayer, the atom configuration is identical to that of an untwisted bilayer (the lattice mismatch between WSe$_2$ and MoSe$_2$ is less than 0.2\%), where one layer is laterally shifted relative to the other layer by a corresponding displacement vector ${\bm d}_0$. For this reason, the moir\'e band structures of twisted TMD bilayers can be constructed from a family of untwisted bilayers at various ${\bm d}_0$, all having $1\times 1$ unit cell. Our analysis thus starts from untwisted bilayers.

In particular, ${\bm d}_0=0, \left(-{\bm a}_{1}+{\bm a}_{2}\right) /3, \left({\bm a}_{1}+{\bm a}_{2}\right) /3$, where ${\bm a}_{1,2}$ is the primitive lattice vector  for untwisted bilayers, correspond to three high-symmetry stacking configurations of untwisted TMD bilayers, which we refer to as MM, XM, MX. In MM (MX) stacking, the M atom on the top layer is locally aligned with the M (X) atom on the bottom layer, likewise for XM. The bilayer structure in these stacking configurations is invariant under three-fold rotation around the $z$ axis.

\begin{figure}[h!]
    \centering
    \includegraphics[width=0.45\textwidth]{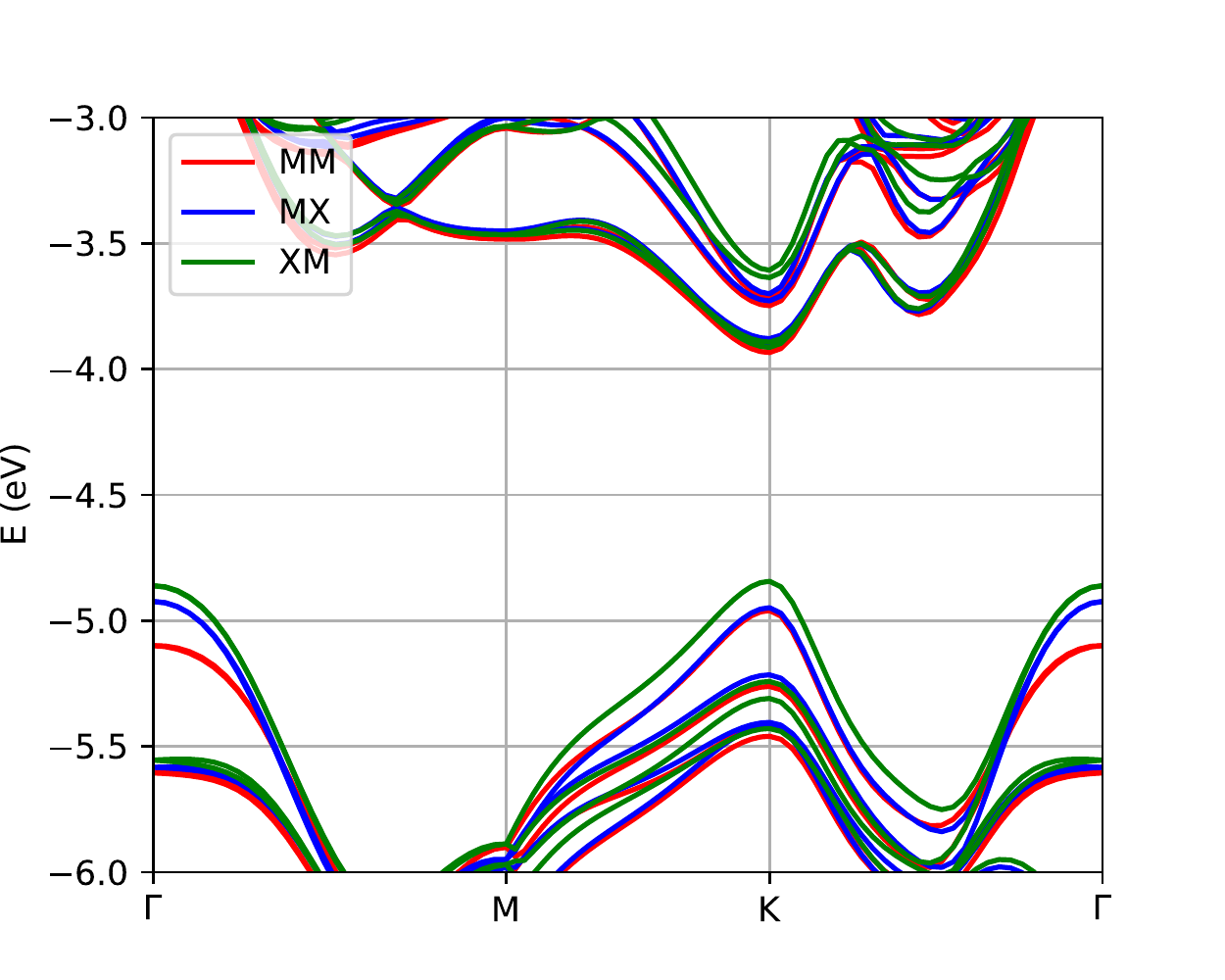}
    \caption{\textbf{DFT band structure for high symmetry stacking structures.}
}
    \label{fig:DFTFig}
\end{figure}

Density functional calculations are performed using generalized gradient approximation with SCAN+rVV10 van der Waals density functional \cite{peng2016versatile}, as implemented in the Vienna ab initio Simulation Package \cite{kresse1996efficient}. Pseudopotentials are used to describe the electron-ion interactions. We first construct the zero-twisted angle WSe$_2$/MoSe$_2$ bilayer at MM, MX, and XM lateral configurations with vacuum spacing larger than 20 \AA{} to avoid artificial interaction between the periodic images along the $z$ direction. The structure relaxation is performed with force on each atom less than 0.001~eV/\AA{}. We use $12\times 12 \times 1$ for structure relaxation and self-consistent calculation. The more accurate SCAN+rVV10 van der Waals density functional  gives the relaxed layer distances as 6.46 \AA{}, 6.45 \AA{} and 6.92 \AA{} for MX, XM and MM stacking structures, respectively. By calculating the work function from electrostatic energy of converged charge density, we obtain the band structure of MM, MX, and XM-stacked bilayers, with reference energy $E=0$ chosen to be the absolute vacuum level, shown in Fig~\ref{fig:DFTFig}.

The low-energy moir\'e band structure for the K valley valence bands is described by the continuum model Hamiltonian~\cite{wu_hubbard_2018},
\begin{equation}
    \mathcal{H}_{\bm{g}',\bm{g}}(\bm{k})\equiv -\frac{\hbar^2 (\bm{k}+\bm{g})^2}{2m^*}\delta_{\bm{g}',\bm{g}}+V\sum_{j=1}^{3}(e^{i\phi}\delta_{\bm{g}',\bm{g}+\bm{g}_j}+e^{-i\phi}\delta_{\bm{g}',\bm{g}-\bm{g}_j}).
\end{equation}
We use $a_M= 13.5$ nm, $m^*=0.5m_e$, and $(V,\phi)=(12.3$ meV,$-125.1^{\circ})$ as obtained from the DFT bands at the three stacking regions. In the limit $\frac{\hbar^2}{2m^*a_M^2}/V\rightarrow \infty$, the model becomes an electron gas and, in the opposite limit, it approaches an array of isolated quantum dots. The long moiré wavelength of our sample $a_M=13.5$ nm puts it closer to the quantum dot array limit with $\frac{\hbar^2}{2m^*a_M^2}/V = 0.034$.
\par The lowest several non-interacting moiré bands in our sample originate from two-dimensional harmonic oscillator-like orbitals on a triangular lattice of $XM$ stacking sites. As holes are doped into moir\'e bands, it is possible that their mutual Coulomb repulsion could cause them to localize to other regions of the moiré unit cell, a scenario known as charge transfer \cite{zhangMoirQuantumChemistry2020}. The small value of $\frac{\hbar^2}{2m^*a_M^2}/V$ in our system suggests that holes will remain localized to triangular lattice sites between filling factors $-2<\nu<-1$. To test this, we calculate the self-consistent, unrestricted HF ground state at spin-projected filling factors $\nu_{\uparrow}=-1, \nu_{\downarrow}=0$ and at $\nu_{\uparrow}=\nu_{\downarrow}=-1$. In the HF approximation, the continuum model Hamiltonian becomes \cite{hu2021competing}
\begin{align}
    \begin{split}
        \mathcal{H}_{s'\bm{g}',s\bm{g}}(\bm{k}) &\equiv\delta_{s',s}[-\frac{\hbar^2 (\bm{k}+\bm{g})^2}{2m^*}\delta_{\bm{g}',\bm{g}}+V\sum_{j=1}^{3}(e^{i\phi}\delta_{\bm{g}',\bm{g}+\bm{g}_j}+e^{-i\phi}\delta_{\bm{g}',\bm{g}-\bm{g}_j})]\\
        &+\frac{1}{A}\sum_{\bm{k}',\bm{g}''}[\delta_{s',s}\sum_{s''}V_{s,s''}(\bm{g}'-\bm{g})\rho_{s'',\bm{g}'+\bm{g}'';s'',\bm{g}+\bm{g}''}(\bm{k}')-V_{s',s}(\bm{g}''+\bm{k}'-\bm{k})\rho_{s',\bm{g}'+\bm{g}'';s,\bm{g}+\bm{g}''}(\bm{k}')].
    \end{split}
\end{align}
The first line of this expression contains the single-particle terms and the second line contains the interaction terms. The first interaction term is the local Hartree potential an the second is the Fock non-local exchange potential. Here $s'$ and $s$ are spin indices, $\rho_{s'\bm{g}';s\bm{g}}\equiv \bra{\Psi_{HF}}c^{\dagger}_{s',\bm{k}+\bm{g'}}c_{s,\bm{k}+\bm{g}}\ket{\Psi_{HF}}$ is the single-particle density matrix in the basis of simultaneous $\bm{p}$, $S_z$ eigenstates, and $\ket{\Psi_{HF}}$ is the single-slater-determinant ground state. We perform the calculation with an $9\times 9$ Monkhorst-Pack Brillouin zone mesh and the lowest $13$ shells of reciprocal lattice vectors. We do not explore states with broken translation symmetry which are unlikely at integer filling factors. We fix the $S_z$ eigenvalue of our state at the outset of the calculation so that $\rho_{s',\bm{g}';s,\bm{g}}\propto \delta_{s',s}$. We use the single-gate-screened Coulomb interaction in the image charge approximation, $V_{s',s}(\bm{q})=\frac{ e^2}{2\epsilon\epsilon_0}\frac{(1-e^{-2dq})}{q}$ where $d=30$ nm is the sample-gate distance and $q\equiv |\bm{q}|$. Additionally, we choose $\epsilon=5$ to account for the hexagonal boron nitride dielectric environment of the superlattice.
\par
\begin{figure}[h!]
    \centering
    \includegraphics[width=\textwidth]{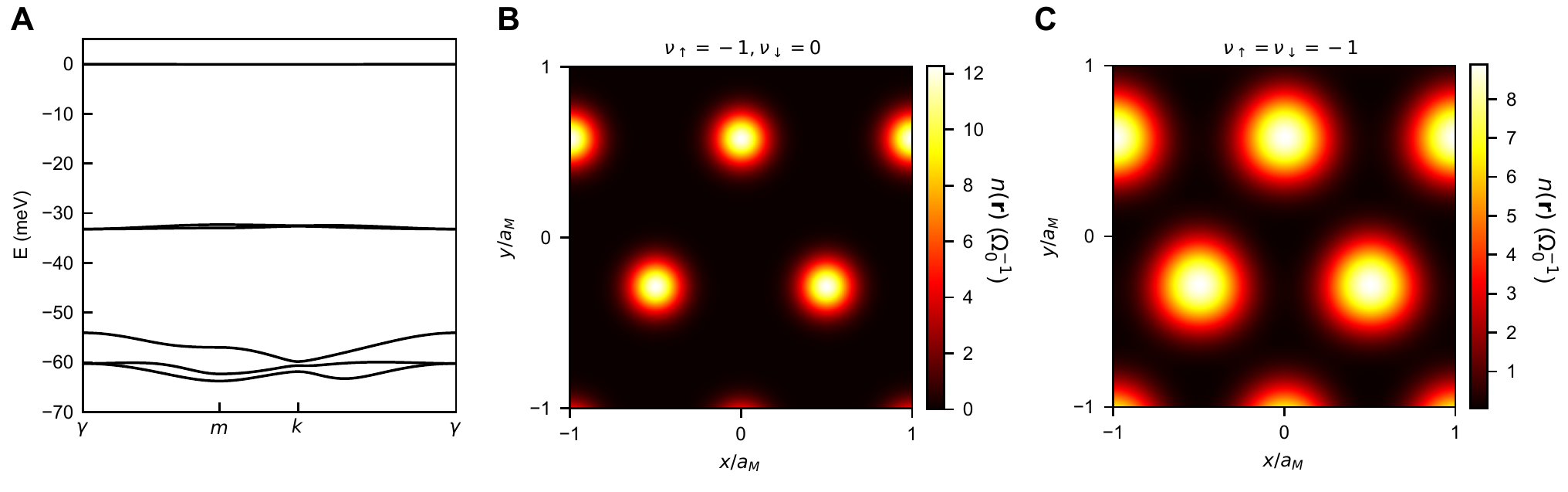}
    \caption{\textbf{Hartree-Fock (HF) study of charge configuration between filling factors $\nu=-1$ and $\nu=-2$.} \textbf{(A)} Non-interacting continuum model band structure. \textbf{(B)} Hole number density of the HF ground state at $\nu_{\uparrow}=-1, \nu_{\downarrow}=0$ and \textbf{(C)} at $\nu_{\uparrow}=\nu_{\downarrow}=-1$. The hole number density in both cases is localized to the $XM$ triangular lattice sites within the moir\'e unit cell.}
    \label{fig:HFFig}
    
\end{figure}

 In Fig. \ref{fig:HFFig}A we plot the non-interacting continuum model band structure. The top (flat) band originates from an $s$ orbital and the next pair of (dispersive) bands originate from two $p$ orbitals at each triangular lattice site. We note that the bandwidth is affected by screening from the underlying filled bands. Hartree terms tend to weaken the moir\'e potential and therefore lead to more delocalized wavefunctions and stronger inter-site orbital hybridization, causing the bandwidth to be increased and filling-dependent. 

In Fig. \ref{fig:HFFig}B, we plot the hole number density of the HF ground state at spin-projected filling factors $\nu_{\uparrow}=-1, \nu_{\downarrow}=0$, and in Fig. \ref{fig:HFFig}C, the same quantity at $\nu_{\uparrow}=\nu_{\downarrow}=1$. In both cases, the charge density within the moiré unit cell is concentrated to the triangular lattice sites of $V_{0}(\bm{r})$ maxima. The charge density within the triangular lattice site at $\nu_{\uparrow}=\nu_{\downarrow}=-1$ is spread out compared to when $\nu_{\uparrow}=-1,\nu_{\downarrow}=0$ because of the repulsion between the two holes on each site. Although we do not enforce $\psi_{n\uparrow{\bm{k}}}=\psi_{n\downarrow{\bm{k}}}$ (in other words, our HF calculation is unrestricted), our self-consistent solution in this case satisfies this condition. These results serve as additional evidence that our system is appropriately modeled as a triangular lattice and that the charge transfer scenario does not occur at least for filling factors $\nu\geq -2$.

Finally, we comment on the possible role of $\Gamma$ valley bands.  At the $MX$ and $XM$ regions, the DFT band structure (Fig.~\ref{fig:DFTFig}) suggests that the $\Gamma$ pocket is competitive in energy with the $K$ pocket.  
While a continuum model describing both $K$ and $\Gamma$ valleys show that the first band is indeed from $K$ valley, the first $\Gamma$ valley band is only $\sim 10$ meV away.
We note that the competition of $\Gamma$ and $K$ valley bands in DFT is very delicate, and depends sensitively on details such as choice of van der Waal density functional and relaxed lattice structure.
In our analyses, we have assumed that the $K$ valley bands are energetically favored and the $\Gamma$ band can be safely ignored.  This is consistent with prior reports that have shown the valence band extrema reside at the $K$ valley for AA stacking configuration \cite{wilson_determination_2017,seyler_signatures_2019,tran_evidence_2019,barre_optical_2022}. We cannot conclusively rule out the possibility that the $\Gamma$ bands are energetically favored in our experiment, but we remark that the underlying physical picture would remain largely unchanged.  
The valley structure can be probed directly in future experiments via magnetic circular dichroism or through the application of an in-plane magnetic field.

For completeness, we additionally remark that an alternative (although less likely) scenario for the origin of the dispersive and flat bands we observe could arise from competition between the $\Gamma$ and $K$ valleys.  
In this scenario, the $\Gamma$ valley bands, which have larger effective mass and feel a stronger moir\'e potential, play the role of the flat bands, while the $K$ valley bands play the role of the dispersive bands.  
However, this scenario requires fine tuning (the first $\Gamma$ and $K$ bands must be very close in energy) and fails to explain some observations. 
First, the presence of observable LL gaps in the dispersive band is at odds with the flatness of the first $K$ band (without significant interaction broadening).
Indeed, in other moir\'e TMD systems with smaller moir\'e period, charge order is generally favored even though the smaller moir\'e unit cell corresponds to larger bandwidth relative to interaction energy.
Additionally, the first $K$ band, associated with an $s$ orbital, should display more prominent hole-like LL gaps (see Fig.~\ref{fig:si_fig_hof}), which is at odds with the observation of only electron-like LL gaps in the first dispersive band.

\section{Phenomenological Stoner model}
The set of parameters used in the phenomenological model to generate Fig.~2D in the main text are as follows: density of states of the dispersive band $D = 1/9t$, effective interaction $U_{fd} = 1/3D = 3t$, band separation $\Delta = 1/D$ (see Fig.~2C), and $g = 0.22W/$Tesla. Here, $t$ is the hopping matrix element of the $p$-orbitals in the tight-binding Hamiltonian and is an adjustable parameter. Parameters are chosen such that the model quantitatively matches the experimental phase diagram in Fig.~1B, but we note that a realistic value of $t \approx 0.2$ meV yields a $g$-factor of ~8$\mu_B$, which is of similar magnitude to values from other measurements in the same system \cite{seyler_signatures_2019}. 

In this model, the width of the reentrant region in which both flat and dispersive bands are partially filled is constant and given by $\delta\nu = (2D U_{fd}-1)/(DU_{fd}-1)$.
In the experimental phase diagram, the width of this region is also roughly constant $\delta\nu\approx 1/2$. This implies that the effective interaction parameter must be $U_{fd}\approx 1/(3D)$. Note, however, that this does not imply that the on-site Hubbard interaction is $1/(3D)$, as the parameter $U_{fd}$ is generically not equal (or even proportional) to the Hubbard interaction. We substantiate this statement through our analysis of an integrable microscopic model in the next section. The model also predicts that for smaller angle (larger $D$), the crossover from filling the flat band to filling the dispersive band should occur at smaller $B$ for given $(\nu_f, \nu_d)$. This is also consistent with our experimental observations in Fig.~\ref{fig:Fig2}. 

\begin{figure}[h!]
    \centering
    \includegraphics[scale =1.0]{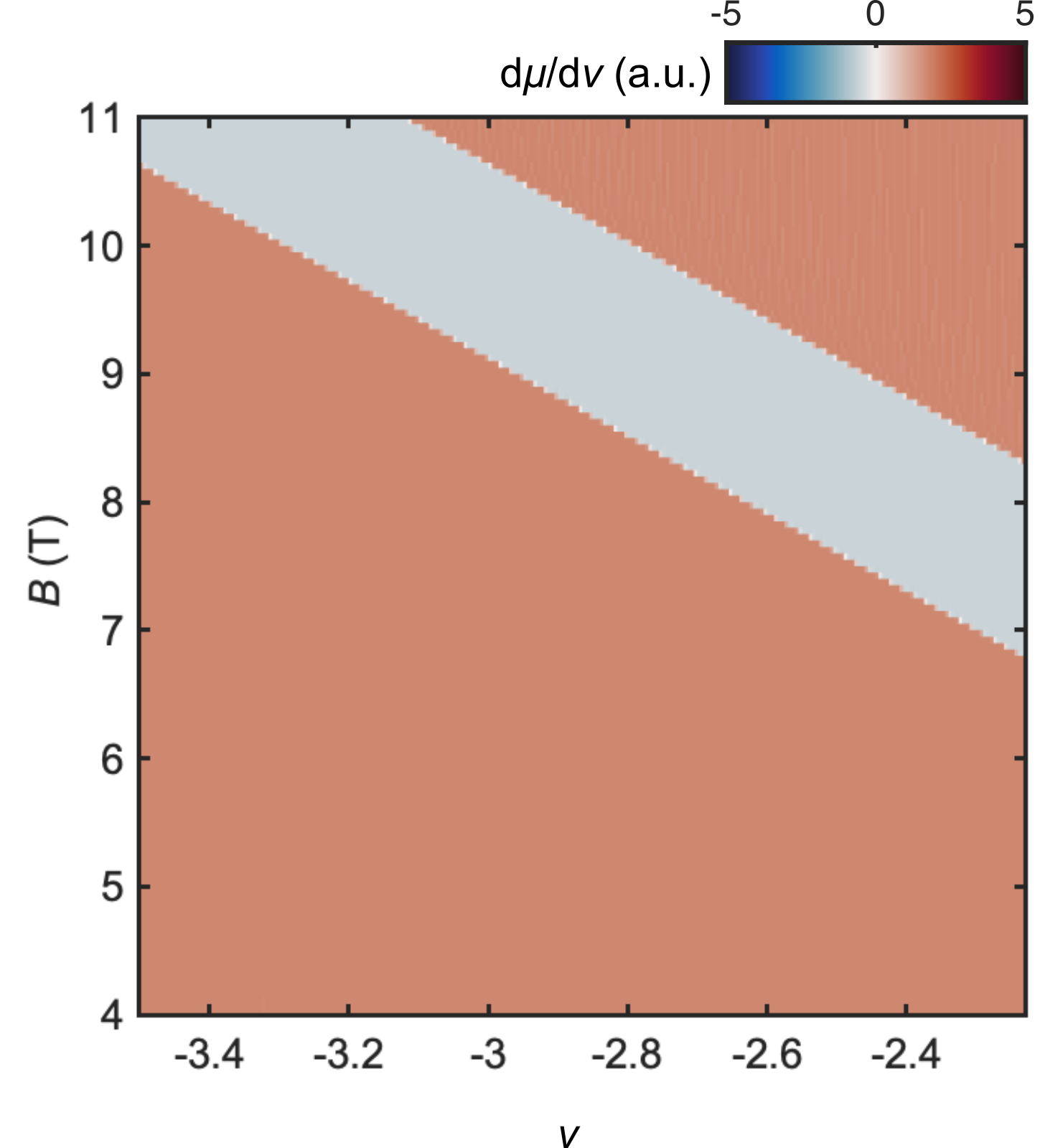}
    \caption{\textbf{Phenomenological Stoner model.} $\dudn$ calculated from the phenomenological Stoner model.} 
    \label{fig:si_fig_pheno}
\end{figure}

Compressibility $\frac{\textrm{d}\mu}{\textrm{d}\nu} = \frac{\textrm{d}^2 E}{{\textrm{d} \nu}^2}$ can be calculated from the phenomenological model, as is shown in Fig. \ref{fig:si_fig_pheno}. Within the region where reentrant charge-ordered states appear, a constant negative $d\mu/d\nu = \frac{-DU_{fd}^2}{1-2DU_{fd}} < 0$ is obtained. Experimentally, the negative compressibility within this region relative to that associated with the adjacent dispersive bands is evident at base temperature in Fig. 4A as well as in Fig. \ref{fig:Fig2}(A,B). The charge-ordered states within this reentrant region consist of states originating from both flat and dispersive bands. The degree to which these distinct states hybridize is an interesting open question to be explored. If the bands do not hybridize, the charge-ordered states would involve simultaneous localization of carriers from both flat and dispersive bands. Alternatively, it is possible that the bands hybridize and produce composite low-energy states with large effective mass, i.e. heavy fermions. 
Finally, we comment on the different behavior we observe while filling the same band at high field between $-2<\nu<-1$ and at low field between $-3<\nu<-2$ (Fig. 1B). This difference may be caused by screening from the underlying filled bands. As noted above, this broadens the dispersion due to Hartree screening. It may also enhance effective Coulomb screening and therefore weaken electronic interactions.

\section{Hubbard model}

We analyze a Hubbard model description for the experimental system in the hole density range $-2\leq \nu \leq -1$ in a magnetic field.
In this region, we assume the the spin-up $s$-orbital flat band is fully filled due to the spin Zeeman effect, and the relevant degrees of freedom are the spin-down $s$-orbital band and dispersive $p$-orbital bands, whose competition is tuned by the magnetic field.
The presence of an insulating state at $\nu=-2$ suggests that only a single $p$ band (the one derived from the spin-up $p$ state with orbital angular momentum quantum number $\ell_z=+1$) is being filled.

We model this system as a two-orbital tight binding model on a triangular lattice, where the two orbitals correspond to the flat and dispersive bands, respectively.
To obtain a tractable model with interactions, we further take the limit where the flat band is perfectly flat and the interaction is only on-site via a Hubbard term.
The resulting Hamiltonian is
\begin{equation}
\mathcal{H} = -t\sum_{\langle i,j\rangle} d^\dagger_i d_j + (\Delta+3|t|-gB)\sum_in^d_i + U\sum_i n^d_i n^{f}_i \label{eq:Hmicroscopic}
\end{equation} 
where $d^\dagger_i, d_i$ are the fermionic creation and annihilation operators for the dispersive band on site $i$, $n^d_i=d^\dagger_i d_i$, and the sum $\langle i,j\rangle$ is over nearest neighbor sites on the triangular lattice.
Finally, $n^f_i=0,1$ is the density of the upper flat band fermions on site $i$ which, due to the absence of a kinetic term, is a good quantum number. 
The hopping parameter $t$ represents the kinetic energy of the dispersive band, and is negative for the $p$ orbital band in question.
The magnetic field is modeled as a Zeeman term $gB$.
We have chosen the constant potential offset $(\Delta+3|t|-gB)$ so that the band bottom is at $\Delta-gB$ in order to make a direct comparison with the phenomenological Stoner model.

The on-site repulsion $U$ is estimated to be much larger than $t$. 
Taking a harmonic approximation for the moir\'e potential with a moir\'e period of $\approx$ 14 nm and effective mass $m^*=0.5m_e$, leads to an effective harmonic oscillator lengthscale $\lambda\approx2$ nm, and corresponding Hubbard energy scale $U\approx \frac{e^2}{4\pi\epsilon\epsilon_0 \lambda} \approx (700/\epsilon)$ meV, where $\epsilon$ is the relative dielectric constant.  
Meanwhile, the non-interacting bandwidth of the $p$ band from the continuum model is estimated to be only $W=9|t|\approx 2$ meV.  
Although the bandwidth will be renormalized due to Coulomb interaction with the filled $s$ band, it is unlikely to change the fact that the local Hubbard interaction is much larger than $t$.  
Thus, we take the limit $U/t \gg 1$, in which there remains only a single energy scale $|t|$ for total density $n^f +  n^d \leq 1$, where $n^{f,d} = |\nu_{f,d}|$.
In this limit, double occupancy is forbidden.
As there is no double occupancy, the interaction energy between the heavy and light bands is purely kinetic: 
the sites on which $n^f_i=1$ act as `defects' for which the dispersive fermions must avoid, thus paying the cost in kinetic energy.

The exact ground state(s) of the above model at a fixed $ n^f$ and $ n^d$ correspond to states with a particular configuration of $n^f_i$ which minimizes the kinetic energy of the dispersive band.  
In reality, however, this effect will be overshadowed by long range Coulomb interactions, which will play a significant role in determining the final charge configuration leading, among other things, to the formation of incompressible generalized Wigner crystal states at commensurate fillings.
For our current analysis, we take a middle ground and simply average over many configurations of $n^f_i$ with equal probability.

We numerically simulate this Hamiltonian on a $L\times L$ triangular lattice torus.  
To do this, we first fix particle numbers $N^f,N^d = 1\dots L^2$, with $N^f+N^d\leq L^2$.
We then solve for the ground state energy of $\mathcal{H}$ with $(\Delta+3|t|-gB)=0$ for $N_{\textrm{samp}}$ random configurations of $n^f_i\in\{0,1\}$, such that $\sum_i n^f_i=N^f$.
The ground state energy for each sample is obtained by diagonalizing the quadratic Hamiltonian for $d$ fermions and filling $N^d$ lowest energy bands.
To decrease finite size effects, we
also average over fluxes threaded through the torus.
The final result is the configuration-averaged energy per unit cell $E_0(n^d,n^f)$, where $n^{(f,d)}=N^{(f,d)}/L^2$.
Note that there is a single energy scale set by $W=9|t|$ in the large $U/t$ limit.  
For a given total density $n^{tot}=n^d+n^f$ and magnetic field $B$, the optimal density distribution can then be obtained by minimizing
$E(n^{tot},B) = \min_{n^d\in[0,n^{tot}]}\left\{ E_0(n^d, n^{tot}-n^d) + (\Delta+3|t|-gB)n^d\right\}$.
Choosing $\Delta=W$ and $g=0.22W/$Tesla as in the Stoner model results in the phase diagram shown in Fig.~2F of the main text.  

\begin{figure}[h!]
    \centering
    \includegraphics[scale =1.0]{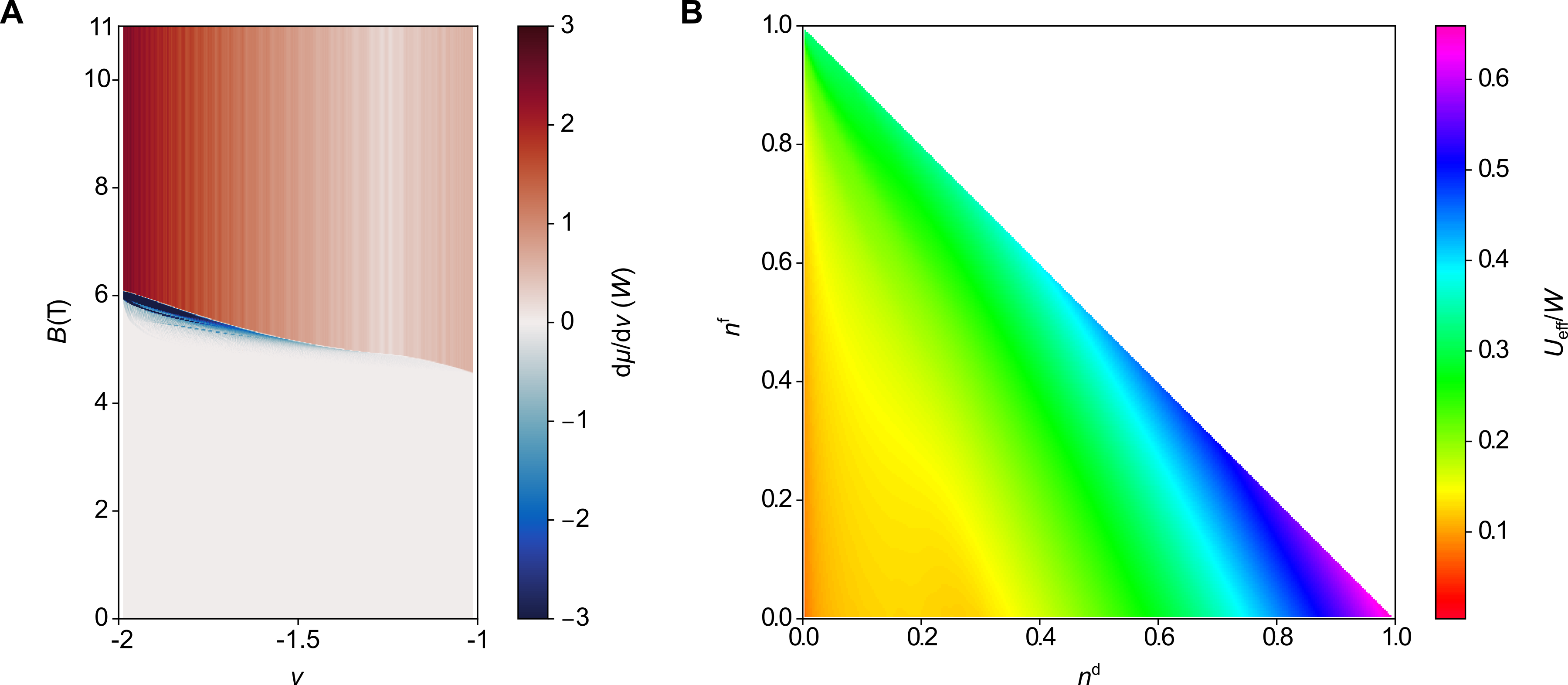}
    \caption{\textbf{Filling-dependent interaction strength and compressibility in the Hubbard model.} \textbf{(A)} Effective interaction strength $U_{eff}$ calculated as a function of flat and dispersive band filling $n^{f,d}$. \textbf{(B)} Numerically evaluated compressibility within the Hubbard model.} 
    \label{fig:si_fig_micro}
\end{figure}

The final result is a relatively sharp transition at a critical field $B_c\approx 6$ T between filling of the flat and dispersive bands.
The critical field varies little with filling, except for a small upturn near $\nu=-1-n^{tot}=-2$.  
Near this upturn, compressibility becomes very negative, as shown in Fig~\ref{fig:si_fig_micro}A, in excellent qualitative agreement with the experimental compressibility measurements in this range.

Finally, we can obtain the effective interaction $U_{fd}$ as a function of $n^f$ and $n^d$ in our effective model.  The energy function $E_0$ can be decomposed as
$E_0(n^d,n^f)=E_0(n^d,0)+E_0(0,n^f)+U_{\mathrm{eff}}(n^d, n^f) n^d n^f$.
The numerically obtained $U_{\mathrm{eff}}(n^d,n^f)/W$ as a function of the dispersive and flat band density is shown in Fig~\ref{fig:si_fig_micro}B.
The value of $U_{\mathrm{eff}}/W\approx 0.3$ in a large part of the diagram, which is consistent with the $U_{fd}=W/3$ suggested by the Stoner model in combination with the experimental width $\delta\nu=1/2$ of the reentrant region (despite this model not being directly applicable for $\nu<-2$).

\section{Singlet-triplet quantum dot transition}
By expanding the moiré potential near its maxima to leading (quadratic) order, we approximate the effective Hamiltonian for a hole isolated to a single moiré potential maximum, as that of a circular harmonic oscillator. The effective Hamiltonian for two Coulomb-interacting holes isolated to the same potential well in a magnetic field is then
\begin{align}
\begin{split}
    H_{QD}&=\hbar\Omega_+(a^{\dagger}_{{1+}}a_{1+}+a^{\dagger}_{{2+}}a_{2+}+1)+\hbar\Omega_-(a^{\dagger}_{{1-}}a_{1-}+a^{\dagger}_{{2-}}a_{2-}+1) +\frac{e^2}{4\pi\epsilon\epsilon_0|\bm{r}_1-\bm{r}_2|}-g\mu_BBS_z.
\end{split}
\end{align}
Here $a_{\alpha\pm}\equiv(a_{\alpha x}\mp ia_{\alpha y})/\sqrt{2}$ lowers the energy of particle $\alpha$ by $\Omega_{\pm}$ and changes its $L_{z}$ eigenvalue by $\pm\hbar$. $\Omega_{\pm}=\sqrt{\omega^2+(\frac{\omega_c}{2})^2}\pm\frac{\omega_c}{2}$, $\omega_c\equiv eB/m^*$ is the effective cyclotron frequency, and $\omega\equiv \sqrt{\frac{16\pi^2V\cos{\phi}}{a_M^2m^*}}$ is the natural harmonic oscillator frequency determined the moiré potential. $g$ is the effective spin $g$-factor. From the continuum model parameters ($V=12.3$ meV, $m^*=0.5m_e$, $\phi=-125.1^{\circ}$, $a_M=13.5$ nm) we find $\hbar\omega =40.3$ meV and $l=1.95$ nm. 
\par The singlet ground state has spin and orbital angular momentum quantum numbers $(S,L)=(0,0)$ and the triplet $(S,L)=(1,1)$. The magnetic field couples to the two particle state through linear and quadratic terms in $\hbar\omega_c$ appearing in $\Omega_{+-}$ and through Zeeman coupling. In the relevant parameter regime, $\omega_c\ll\omega_0$. For this reason, the spin singlet state is decoupled from the magnetic field [with leading corrections of $\mathcal{O}(\omega_c/\omega)^2$] and the triplet state is coupled linearly to the magnetic field through its orbital and spin angular momentum. In the absence of the Coulomb interaction, the singlet-triplet gap is $\Delta_{st}(B)\equiv E_{0}(S=1)-E_0(S=0)=\hbar\Omega_{-}-g\mu_BB$. The singlet-triplet ground state transition occurs at $B_{c}$ such that $\Delta_{st}(B_c)=0$. As the strength of the Coulomb interaction increases, $\Delta_{st}(B)$ monotonically decreases for all $B$ and thus so too does $B_c$.
\par To determine $B_c$ in the presence of a finite Coulomb interaction, we solve this two-body problem using exact diagonalization following the approach of \cite{zeng2022strong}, to which we refer the reader for a technical discussion of the calculation. We plot the critical magnetic field for the singlet-triplet transition $B_c(\epsilon^{-1})$ as a function of Coulomb interaction strength for several values of $g$. In the absence of the Coulomb interaction, $B_c\gg6$ T where $6$ T is the approximate value of the observed phase transition in the filling range $-2>\nu>-1$. In the presence of a finite Coulomb interaction, however, $B_c$ monotonically decreases, approaching values comparable to $6$ T. We note that this singlet-triplet transition has been studied theoretically and observed experimentally in GaAs quantum dots \cite{wagner1992spin, ashoori1993n}.
\par Finally, we comment on our observation that $B_c$ is smaller in regions of smaller twist angle. This is consistent with the predicted scaling of $B_c$ with twist angle in our quantum dot model. Since, as noted above, $\omega_c\ll \omega_0$, $\Delta_{st}(B)$ is a linear function of $B$. In the absence of Coulomb interactions, $\Delta_{st}(B=0)=\hbar\omega\propto a_M^{-1} \propto \theta$. The interaction strength is characterized by the dimensionless ratio of energy scales $\frac{e^2/(4\pi\epsilon\epsilon_0 l)}{\hbar\omega} \propto a_M^{1/2}\propto \theta^{-1/2}$. As a function of interaction strength, $\Delta_{st}$ decreases monotonically. Since the non-interacting gap decreases with decreasing $\theta$ and the coupling constant, which decreases the gap, increases with decreasing $\theta$, $\Delta_{st}(B=0)$ and therefore also $B_c$ monotonically decrease with decreasing $\theta$.
\begin{figure}[h!]
    \centering
    \includegraphics{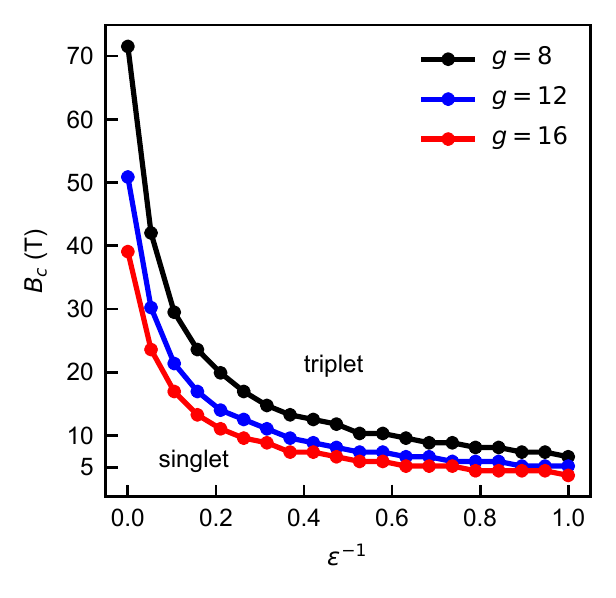}
    \caption{\textbf{Critical magnetic field for singlet-triplet transition of the two-hole quantum dot.} Value of the applied magnetic field at which the two-hole ground state within a moiré potential well undergoes a singlet-triplet transition as a function of Coulomb interaction strength $\epsilon^{-1}$ determined by exact diagonalization of two Coulomb-interacting holes in a circular harmonic potential.}
    \label{fig:singletTripletBc}
\end{figure}
\section{Single-particle Hofstadter spectrum}
\begin{figure}[h!]
    \centering
    \includegraphics{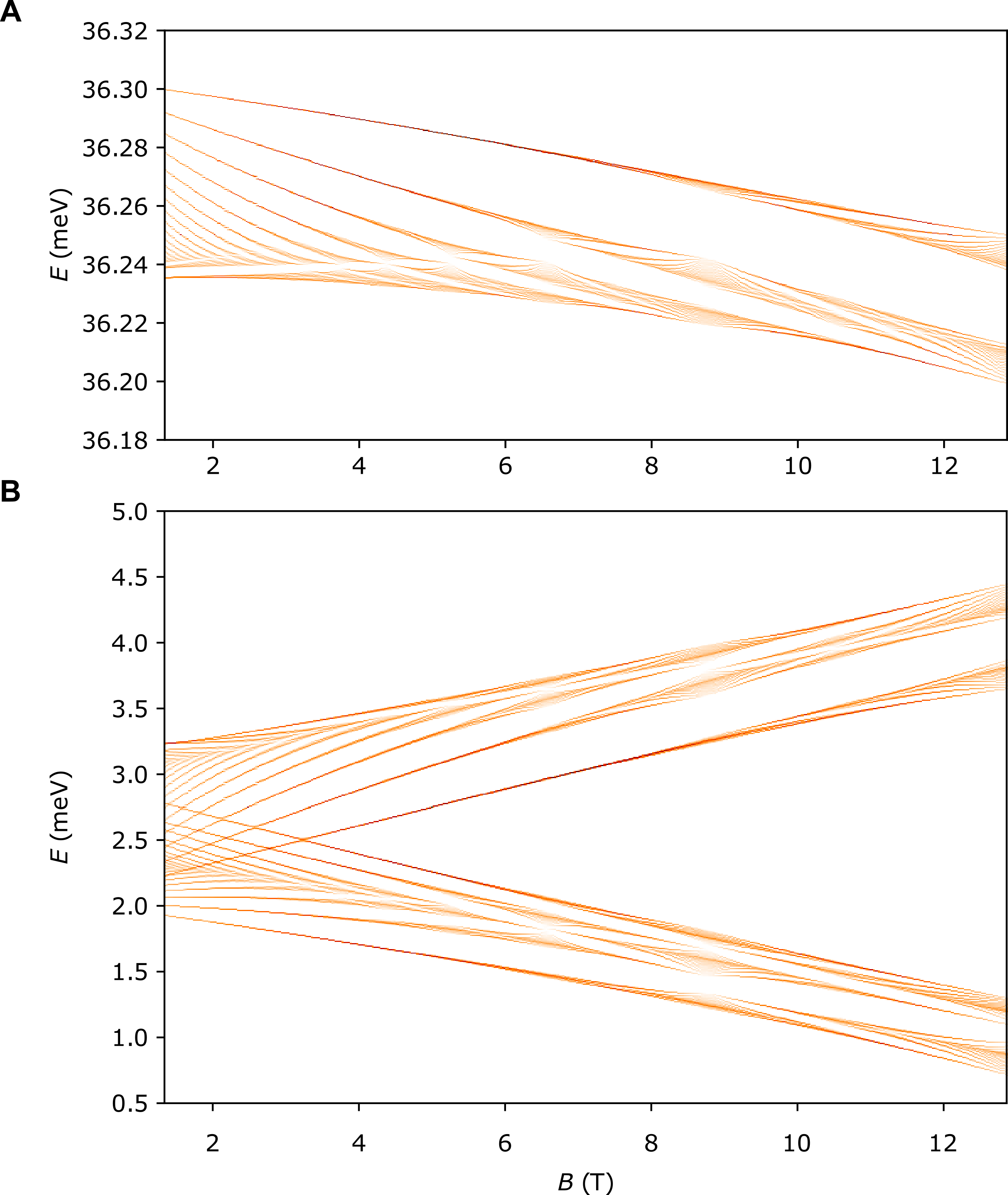}
    \caption{\textbf{Single-particle Hofstadter spectrum.} \textbf{(A-B)} Single-particle and spinless Hofstadter spectrum calculated from the continuum model for the flat $s$ (A) and dispersive $p$ (B) bands.}
    \label{fig:si_fig_hof}
\end{figure}

In Fig.~\ref{fig:si_fig_hof}, we show the single-particle Hofstadter spectrum for the first three bands of a single spin species, computed from the continuum model.
We work with the continuum model,
\begin{equation}
H(\mathbf{k}) = -\frac{\mathbf{k}^2}{2m} + V(\mathbf{r})
\end{equation}
where $V(\mathbf{r}) = 2v\sum_{i=1,3,5}\cos(\mathbf{g}_i\cdot \mathbf{r}+\phi)$.
A magnetic field is introduced via minimal coupling by substituting $\mathbf{k}\rightarrow\bm{\pi}\equiv \mathbf{k}-\mathbf{A}$, where $\mathbf{A}=(B/2)(x \hat{\mathbf{y}} - y\hat{\mathbf{x}})$ is the vector potential in the symmetric gauge.
The gauge-invariant operators $\pi_x$ and $\pi_y$ satisfy the commutation relation $[\pi_x,\pi_y]=iB$.  
From these, we define Landau level raising  operator $a^\dagger = (\pi_x-i\pi_y)/\sqrt{2B}$.
We also define the operators $\mathbf{K}=\mathbf{k}+\mathbf{A}$, which are gauge-dependent and satisfy $[\pi_i,K_i]=0$ and $[K_x,K_y]=-iB$.
We work in the eigenbasis of $n=a^\dagger a$ and $K_x$, spanned by the states $|n,K_x\rangle$ where $n=0,1,2,\dots$ and $K_x\in\mathbb{R}$.

The kinetic energy is diagonal in this basis is independent of $K_x$
\begin{equation}
-\frac{\bm{\pi}^2}{2m}\ket{n,K_x} = -\frac{B}{m}\left(n+\frac{1}{2}\right)\ket{n,K_x}
\end{equation}
The potential term can be decomposed into terms of the form $e^{i\mathbf{Q}\cdot \mathbf{r}}$,  which act in this basis as
\begin{equation}
e^{i\mathbf{Q}\cdot\mathbf{r}}\ket{n,K_x} = \sum_{n^\prime} D_{n^\prime n}(\frac{Q_x+iQ_y}{\sqrt{2B}})  e^{-i \frac{Q_y}{B}\left(K_x+\frac{Q_x}{2}\right)}\ket{n^\prime,K_x+Q_x}
\end{equation}
where 
\begin{equation}
D_{n^\prime n}(z) = 
\begin{cases}
z^{n^\prime-n}e^{-\frac{|z|^2}{2}} \sqrt{\frac{n^\prime !}{n !}}L^{(n^\prime-n)}_n(|z|^2) & n^\prime \geq n \\
 (-z^*)^{n-n^\prime}e^{-\frac{|z|^2}{2}}\sqrt{\frac{n !}{n^\prime !}}L^{(n-n^\prime)}_{n^\prime}(|z|^2) & n^\prime < n 
\end{cases}
\end{equation}

The reciprocal lattice vectors can be expressed as 
\begin{equation}
\mathbf{g}_1 = (0,2\Delta_y);\;\;
\mathbf{g}_3 = (-\Delta_x,-\Delta_y);\;\;
\mathbf{g}_5 = (\Delta_x,-\Delta_y)
\end{equation}
where $\Delta_x=\sqrt{3}g/2,\Delta_y=g/2$, and $g=4\pi/(\sqrt{3}a_M)$.
In terms of $\Delta_{x,y}$, it becomes clear that the potential term only mixes $K_x$ with $K_x\pm\Delta_x$, and is periodic under $K_x\rightarrow K_x+p\Delta_x$ if $B=p\Delta_y\Delta_x/(2\pi q)$, for integers $p,q$.
We therefore introduce a new basis
\begin{equation}
|n,j;x_0,k_0\rangle = \mathcal{N}\sum_{\ell\in\mathbb{Z}} e^{\frac{2\pi i k_0}{p} (x_0+j+\ell p)}|n,K_x=\Delta_x(x_0+j+\ell p)\rangle
\end{equation}
up to a normalization factor, where $x_0\in[0,1]$ and $k_0\in[0,1]$, and $j=1,\dots,p$ with $j\equiv j+p$.
Each term in the potential $e^{i\mathbf{Q}\cdot\mathbf{r}}$ can be characterized by two integers $q_x = Q_x/\Delta_x$ and $q_y=Q_y/\Delta_y$, which act in this basis as
\begin{equation}
e^{i\mathbf{Q}\cdot\mathbf{r}}|n,j;x_0,k_0\rangle = \sum_{n^\prime}D_{n^\prime n}(\frac{Q_x+iQ_y}{\sqrt{2B}})e^{- \frac{2\pi i q }{p} q_y(x_0+j+\frac{q_x}{2})-2\pi i k_0 \frac{q_x}{p}}\ket{n,j+q_x;x_0,k_0}
\end{equation}

In practice, we truncate the Landau level basis to some large but finite number $n=0,\dots,N_{\mathrm{max}}$, which must be chosen to not affect the low energy eigenvalues of interest.
For each fixed $x_0$, $k_0$, the resulting Hamiltonian is $p N_{\mathrm{max}}\times p N_{\mathrm{max}}$ dimensional and can be exactly diagonalized to obtain a set of eigenvalues $E^{x_0 k_0}_i$.
The density of states is obtained by averaging over $x_0$ and $k_0$ with a factor of $1/(2q)$,
\begin{equation}
\mathrm{DOS}(E)=\frac{1}{2q A_{uc}}\left\langle \sum_i \delta(E-E^{x_0 k_0}_i)\right\rangle_{x_0 k_0}
\end{equation}
where $A_{uc}=\sqrt{3}a_M^2/2$ is the moir\'e unit cell area.

The results are shown in Fig.~\ref{fig:si_fig_hof} for the first three moir\'e bands (of a single spin species) with a moir\'e period of $a_M=13.5$ nm, effective mass $m=0.5m_e$, potential strength $v=12.3$ meV and phase factor $\phi=-125.1^\circ$.
For a given $p,q$, we keep $N_{\mathrm{max}}=\lfloor 100q/p\rfloor$ states which is sufficiently large to avoid cutoff effects in the first three bands, and average over randomly chosen $x_0,k_0\in[0,1]$ until a sufficiently smooth DOS is achieved.

The Hofstadter spectra reveals many features which are consistent with experimental observations at low density.   
The first band is derived from the $s$ orbital of the moir\'e site and is exceptionally flat.  
The energy gaps between Landau levels (Hofstadter subbands) within the first band is on the order of $0.01$ meV, much too small to be detected experimentally.  
Indeed, in the density range $-2\leq \nu \leq 0$ for $B<6$ T where the flat band is being filled, no Hofstadter gaps are observed.

The second and third bands are derived from the $p$ orbitals with orbital angular momentum $\ell_z=\pm 1$.  
The magnetic field couples to the orbital angular momentum leading to a linear splitting of the two bands with $B$.  
The lowest energy state after the $s$ orbital bands have been filled is the state with orbital angular momentum $\ell_z=+1$ and spin aligned with $B$.
In the experimental phase diagram, we expect this state to be filled in the region $-3\leq \nu \leq -2$ for $B<6$ T, and $-2\leq \nu \leq -1$ for $B>6$ T due to the interaction induced band reordering.
In both these regions, only electron-like Hofstadter gaps with positive slope are observed, i.e. $(t,s)$ with positive $t$.
This is consistent with the Hofstadter spectrum in Fig.~\ref{fig:si_fig_hof} which shows that only electron-like Landau level gaps in the second band are appreciable.
This can be understood as arising to an asymmetry in the dispersion of the band: the band features a flat band top but dispersive band bottom.  Thus, only Landau levels emanating from the band bottom, i.e. electron-like Landau levels, are observable.

For filling beyond this, we expect the bands to be strongly renormalized by interactions with filled states and no longer resemble that of the single particle band structure. 
The order in which the bands fill will also sensitively depend on interactions and spin $g$-factor. The closing and reopening of gaps at fixed filling $\nu=-3$ and $\-2.5$ are likely also related to changes in the underlying band occupation.
A comprehensive theoretical study of the interaction-renormalized band structure at higher filling, and the corresponding Hofstadter spectrum is beyond the scope of this work.


\section{Gaps of the Hofstadter states}

\begin{figure}[h]
    \centering
    \includegraphics[scale =1.0]{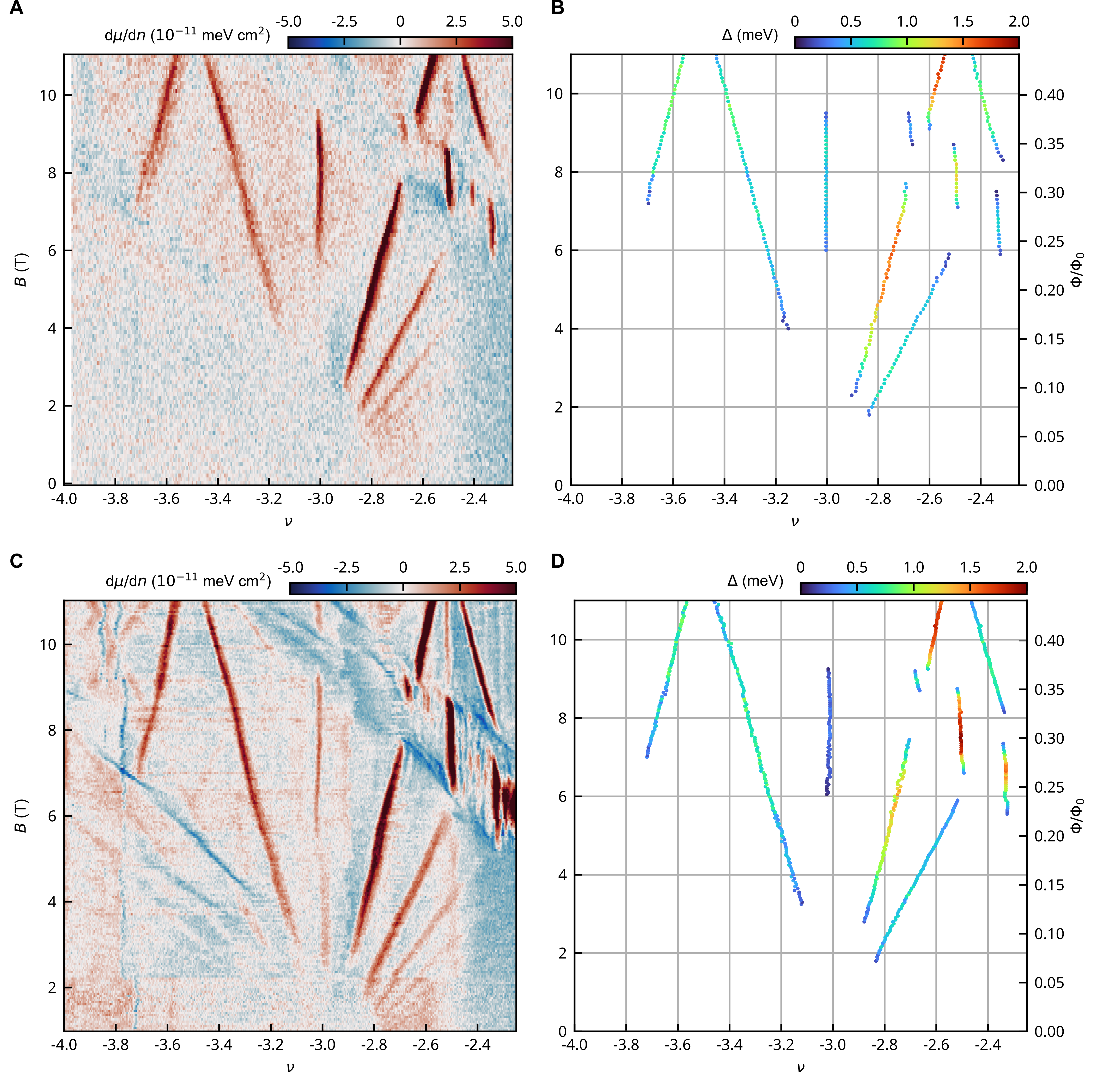}
    \caption{\textbf{Hofstadter gaps for $-4<\nu<-2$.} \textbf{(A,C)} Zoom-ins from Figs.~1B and 4A in the main text, which were measured at two independent locations. \textbf{(B,D)} Gaps extracted from (A) and (C), respectively. (A) was taken at temperature $T=1.6$ K while (C) at $T=330$ mK and with a smaller tip-sample separation.}
    \label{fig:HofstadterWannierGaps_Loc5}
\end{figure}

We extract the Hofstadter gaps from measurements in two different locations (Fig.~\ref{fig:HofstadterWannierGaps_Loc5}A,C) and plot them as Wannier diagrams in Fig.~\ref{fig:HofstadterWannierGaps_Loc5}B,D. Both plots show similar quantitative and qualitative trends with magnetic field. The Hofstadter gaps first increase with field after emerging but then decrease rapidly, both near the region of reentrant charge order and also when approaching $\Phi/\Phi_0= 1/2$. Additionally, the Wannier plot further demonstrates the asymmetry in the second moir\'e band dispersion.  In particular, above the reentrant charge order for $-3<\nu<-2$, we observe larger electron-like than hole-like Hofstadter gaps in agreement with the different effective masses for top and bottom band edges (this is also evident from the size of the Hofstadter gaps for $-2<\nu<-1$ at high fields).

\FloatBarrier
\bibliographystyle{apsrev4-1}
\bibliography{references.bib,references-2.bib}